\input harvmac.tex
\input diagrams 

%
\let\includefigures=\iftrue
%
\let\useblackboard=\iftrue
%
%
\newfam\black
\input epsf
\includefigures
\message{If you do not have epsf.tex (to include figures),}
\message{change the option at the top of the tex file.}
\def\figin{\epsfcheck\figin}\def\figins{\epsfcheck\figins}
\def\epsfcheck{\ifx\epsfbox\UnDeFiNeD
\message{(NO epsf.tex, FIGURES WILL BE IGNORED)}
\gdef\figin##1{\vskip2in}\gdef\figins##1{\hskip.5in}
\else\message{(FIGURES WILL BE INCLUDED)}%
\gdef\figin##1{##1}\gdef\figins##1{##1}\fi}
\def\DefWarn#1{}
\def\figinsert{\goodbreak\midinsert}
\def\ifig#1#2#3{\DefWarn#1\xdef#1{fig.~\the\figno}
\writedef{#1\leftbracket fig.\noexpand~\the\figno}%
\figinsert\figin{\centerline{#3}}\medskip\centerline{\vbox{\baselineskip12pt
\advance\hsize by -1truein\noindent\footnotefont{\bf Fig.~\the\figno:} #2}}
\bigskip\endinsert\global\advance\figno by1}
\else
\def\ifig#1#2#3{\xdef#1{fig.~\the\figno}
\writedef{#1\leftbracket fig.\noexpand~\the\figno}%
\global\advance\figno by1}
\fi

\def\IZ{\relax\ifmmode\mathchoice
{\hbox{\cmss Z\kern-.4em Z}}{\hbox{\cmss Z\kern-.4em Z}}
{\lower.9pt\hbox{\cmsss Z\kern-.4em Z}}
{\lower1.2pt\hbox{\cmsss Z\kern-.4em Z}}\else{\cmss Z\kern-.4em
Z}\fi}
\def\inbar{\,\vrule height1.5ex width.4pt depth0pt}
\def\IB{\relax{\rm I\kern-.18em B}}
\def\IC{\relax\hbox{$\inbar\kern-.3em{\rm C}$}}
\def\ID{\relax{\rm I\kern-.18em D}}
\def\IE{\relax{\rm I\kern-.18em E}}
\def\IF{\relax{\rm I\kern-.18em F}}
\def\IG{\relax\hbox{$\inbar\kern-.3em{\rm G}$}}
\def\IH{\relax{\rm I\kern-.18em H}}
\def\II{\relax{\rm I\kern-.18em I}}
\def\IK{\relax{\rm I\kern-.18em K}}
\def\IL{\relax{\rm I\kern-.18em L}}
\def\IM{\relax{\rm I\kern-.18em M}}
\def\IN{\relax{\rm I\kern-.18em N}}
\def\IO{\relax\hbox{$\inbar\kern-.3em{\rm O}$}}
\def\IP{\relax{\rm I\kern-.18em P}}
\def\IQ{\relax\hbox{$\inbar\kern-.3em{\rm Q}$}}
\def\IR{\relax{\rm I\kern-.18em R}}
\font\cmss=cmss10 \font\cmsss=cmss10 at 7pt
\def\IZ{\relax\ifmmode\mathchoice
{\hbox{\cmss Z\kern-.4em Z}}{\hbox{\cmss Z\kern-.4em Z}}
{\lower.9pt\hbox{\cmsss Z\kern-.4em Z}}
{\lower1.2pt\hbox{\cmsss Z\kern-.4em Z}}\else{\cmss Z\kern-.4em Z}\fi}
\def\IGa{\relax\hbox{${\rm I}\kern-.18em\Gamma$}}
\def\IPi{\relax\hbox{${\rm I}\kern-.18em\Pi$}}
\def\ITh{\relax\hbox{$\inbar\kern-.3em\Theta$}}
\def\IOm{\relax\hbox{$\inbar\kern-3.00pt\Omega$}}

\def\inbar{\,\vrule height1.5ex width.4pt depth0pt}

\font\cmss=cmss10 \font\cmsss=cmss10 at 7pt
\def\IR{\relax{\rm I\kern-.18em R}}

\lref\dm{M.R. Douglas and G. Moore, ``D-branes, quivers, 
and ALE instantons,'' hep-th/9603167.}
\lref\hm {J. Harvey and G. Moore, ``On the algebras of BPS states,'' 
hep-th/9609017, Commun. Math. Phys. {\bf 197} (1998) 489.}
\lref\bpsproc{G. Moore, 
``String duality, automorphic forms, and generalized Kac-Moody algebras,'' 
hep-th/9710198, Nucl. Phys. {\bf B}, Proc. Suppl. {\bf 67} (1998) 56. }
\lref\ds{M. Dine and N. Seiberg ``Non renormalization theorems is superstring
theory,''Phys. Rev. Lett. {\bf 57} (1986) 2625.}
\lref\marti{E. Martinec ``Non renormalization theorems and fermionic string
finiteness,'' Phys. Lett. {\bf B 171} (1986), 189.}
\lref\giveon{A. Giveon and M. Porrati {\it LEEA}}
\lref\kallosh{R. Kallosh and Morozov ``On vanishing of multiloop
contributions to 0,1,2,3 point functions in Green-Schwarz formalism for
heterotic string,'' Phys.Lett. {\bf B 207} (1988) 164.} 
\lref\pistab{M.R. Douglas, B. Fiol and C. R\"omelsberger, ``Stability and BPS
branes,'' hep-th/0002307}
\lref\mmms{M. Mari\~no, R. Minasian, G. Moore and A. Strominger, 
``Nonlinear instantons from supersymmetric $p$-branes,''}
\lref\king{A.D. King, ``Moduli of representations of finite 
dimensional algebras,'' Quart. J. Math. Oxford {\bf 45} (1994) 515. }
\lref\koba{S. Kobayashi, {\it Differential geometry of complex 
vector bundles,} Princeton University Press.}
\lref\sharpe{E. Sharpe, ``K\"ahler cone substructure,'' hep-th/9810064, 
Adv. Theor. Math. Phys. {\bf 2} (1998) 1441. }
\lref\potier{J. Le Potier, {\it Lectures on vector bundles,} Cambridge 
University Press, 1997. }
\lref\sethi{S. Sethi and M. Stern, ``D-brane bound states redux,'' 
hep-th/9705046, Commun. Math. Phys. {\bf 194} (1998) 675.}
\lref\bilal{F. Ferrari and A. Bilal, ``The strong coupling spectrum of 
Seiberg-Witten theory,'' hep-th/9602082, Nucl. Phys. {\bf B 469} (1996) 387.}
\lref\kron{P.B. Kronheimer, ``The construction of ALE spaces as 
hyperK\"ahler quotients,'' J. Differ. Geom. {\bf 28} (1989) 665. }
\lref\kn{P.B. Kronheimer and H. Nakajima, ``Yang-Mills instantons 
on ALE gravitational instantons,'' Math. Ann. {\bf 288} (1990) 263. }
\lref\dfr{M.R. Douglas, B. Fiol and C. R\"omelsberger, ``The spectrum of 
BPS branes on a noncompact Calabi-Yau,'' hep-th/0003263.}
\lref\nakapaper{H. Nakajima, ``Varieties associated with quivers,'' 
in {\it Representation theory of algebras and
related topics} (Mexico City, 1994), 139, CMS Conf. Proc., 19, 
Amer. Math. Soc., Providence, RI, 1996.}
\lref\nakalg{H. Nakajima, ``Instantons on ALE spaces, quiver varieties, 
and Kac-Moody algebras,'' Duke Math. J. {\bf 76} (1994) 365.}
\lref\kactwo{V.G. Kac, ``Infinite root systems, representations of 
graphs and invariant theory,'' Inv. Math. {\bf 56} (1980) 57.}
\lref\md{M.R. Douglas, ``Enhanced gauge symmetry in M(atrix) theory,'' 
hep-th/9612126, JHEP {\bf 9707} (1997) 004. }
\lref\ddg{E. Diaconescu, M.R. Douglas and J. Gomis, 
``Fractional branes and wrapped branes,'' hep-th/9712230, JHEP {\bf 9802} 
(1998) 013 }
\lref\dgm{M.R. Douglas, B.R. Greene and D.R. Morrison, 
``Orbifold resolution by D-branes,'' hep-th/9704151, Nucl. Phys. {\bf B 506} 
(1997) 84. } 
\lref\dg{E. Diaconescu and J. Gomis, ``Duality in Matrix theory and 
three-dimensional mirror symmetry,'' hep-th/9707019, 
Nucl. Phys. {\bf B 517} (1998) 53.}
\lref\sesha {C.S. Seshadri, ``Space of unitary vector bundles on a 
compact Riemann surface,'' Ann. Math. {\bf 85} (1967) 303. }
\lref\dsharpe {E. Sharpe ``D-branes, derived categories, and 
Grothendieck groups,'' hep-th/9902116, Nucl. Phys. {\bf B 561} (1999) 
433. }
\lref\daspin {P. Aspinwall and R. Donagi, ``The heterotic string, 
the tangent bundle, and derived categories,'' hep-th/9806094, 
Adv. Theor. Math. Phys. {\bf 2} (1998) 1041.}
\lref\mirror{M. Kontsevich, ``Homological algebra of mirror symmetry,'' 
alg-geom/9411018.}
\lref\gm {S.I. Gelfand and Y.I. Manin, {\it Methods of homological algebra,} 
Springer-Verlag, 1999.}
\lref\frenkel{I. Frenkel, A. Malkin and M. Vybornov, ``Affine Lie 
algebras and tame quivers,'' math.RT/0005119. }
\lref\hitchin{N. Hitchin, ``$L^2$-cohomology of hyperK\"ahler quotients,'' 
Commun. Math. Phys. {\bf 211} (2000) 153.}
\lref\kacbook{V.G. Kac, {\it Infinite dimensional Lie algebras}, 
Cambridge University Press, 1990. }
\lref\ringel{C.M. Ringel, ``Hall algebras,'' in {\it Topics in algebra,} 
Banach Center Publications, vol. 26, Part I, PWN Polish Scientific 
Publishers, Warsaw 1990; ``Hall algebras and quantum groups,'' 
Inv. Math. {\bf 101} (1990) 583. }
\lref\ringeltwo{C.M. Ringel, ``Hall polynomials for the representation-finite 
hereditary algebras,'' Adv. Math. {\bf 84} (1990) 137.}
\lref\sw{N. Seiberg and E. Witten, ``Electric-magnetic duality, 
monopole condensation, and confinement in ${\cal N}=2$ supersymmetric 
Yang-Mills theory,'' hep-th/9407087, Nucl. Phys. {\bf B 426} (1994) 
19.} 
\lref\gv{R. Gopakumar and C. Vafa, ``M-theory and topological strings, II,'' 
hep-th/9812127.}
\lref\dh{A. Dabholkar and J. Harvey ``Nonrenormalization of the superstring
tension,'' Phys. Rev. Lett. {\bf 478} (1989) 63.}
\lref\kpapers {R. Minasian, G. Moore, ``K-theory and Ramond-Ramond charge,'' 
hep-th/9710230,  JHEP {\bf 9711} (1997) 002. 
E. Witten, ``D-branes and K-theory,'' hep-th/9810188, 
JHEP {\bf 9812} (1998) 019.}
\lref\bbs {K. Becker, M. Becker, and A. Strominger, ``Fivebranes, membranes, 
and nonperturbative string theory,'' hep-th/9507158, Nucl. Phys. 
{\bf B 456} (1995) 130.}
\lref\ooguri{H. Ooguri, Y. Oz, and Z. Yin, ``D-branes on Calabi-Yau 
spaces and their mirrors,'' hep-th/9606112, Nucl. Phys. {\bf B 477} (1996) 
407.}
\lref\qpapers {D. J. Benson, {\it Representations and Cohomology I \& II} 
Cambridge University Press, 30,31, Cambridge (1991). H. Kraft and Ch. 
Riedtmann, ``Geometry of representations of quivers,'' in 
{\it Representations of 
algebras,} Proceedings of the Durham Symposium 1985, LMSLNS {\bf 116}, 
Cambridge University Press, 1986.}
\lref\dmike {M.R. Douglas, ``D-branes and categories,'' to appear.}
\lref\rusos{A. Ritz, M. Shifman, A. Vainshtein and M. Volosin, 
``Marginal stability and the metamorphosis of BPS states,'' 
hep-th/0006028.}
\lref\catop{J.J. Atick, G. Moore, and A. Sen ``Catoptric tadpoles,'' Nucl. Phys.
{\bf B 307} (1988) 221.}
\lref\phases {E. Witten, ``Phase transitions in 
M-theory and F-theory,'' hep-th/9603150, Nucl. Phys. {\bf B 471} (1996) 
195.}
\lref\ceco {S. Cecotti and C. Vafa, ``On classification of ${\cal N}=2$ 
supersymmetric theories,'' hep-th/9211097, Commun. Math. Phys. 
{\bf 157} (1993) 139. }
\lref\aspin {P. Aspinwall, ``Enhanced gauge symmetries and K3 surfaces,'' 
hep-th/9507012, Phys. Lett. {\bf B 357} (1995) 329.}
\lref\ddm {D.-E. Diaconescu and M.R. Douglas, ``D-branes on 
stringy Calabi-Yau manifolds,'' hep-th/0006224.}
\lref\sd{E. Witten, ``String theory dynamics in various dimensions,'' 
hep-th/9503214, Nucl. Phys. {\bf B 443} (1995) 85.}
\lref\mirrorvafa{K. Hori and C. Vafa, ``Mirror symmetry,'' hep-th/0002222. }
\lref\mirrord{K. Hori, A. Iqbal, and C. Vafa, ``D-branes and 
mirror symmetry,'' hep-th/0005247.}
\lref\chiral{W. Lerche, C. Vafa and N. Warner, ``Chiral rings in ${\cal N}=2$ 
superconformal theories,'' Nucl. Phys. {\bf B 324} (1989) 427.}
\lref\verlinde {E. Verlinde and H. Verlinde, ``Multiloop 
calculations in covariant
superstring theory,'' Phys.Lett. {\bf B 192} (1987) 95.}
\lref\shenker {S.H. Shenker, ``The strength of non-perturbative effects in
string theory,'' in Carg\`ese Workshop on Random Surfaces, 
Quantum Gravity and Strings, 1990.}
\lref\eva {E. Silverstein , ``Duality, Compactification, and 
$e^{-1/\lambda}$ Effects in the Heterotic String Theory,'' Phys. Lett. 
{\bf B 396} (1997) 91.}
\lref\boris {I. Antoniadis, B. Pioline, T.R. Taylor, {\it Calculable 
e^{-1/\lambda} Effects}, Nucl.Phys. B512 (1998) 61-78}
\lref\dtopics{M.R. Douglas, ``Topics in D-geometry,'' hep-th/9910170, 
Class. Quant. Grav. {\bf 17} (2000) 1057.}
\lref\shamit{S. Kachru, S. Katz, A. Lawrence and J. McGreevy, ``Open string
instantons and superpotentials,'' hep-th/9912151, 
Phys. Rev. {\bf D 62} (2000) 026001; ``Mirror symmetry for 
open strings,'' hep-th/0006047}
\lref\govin{S. Govindarajan and T. Jayaraman, ``On the Landau-Ginzburg 
description of boundary CFTs and special Lagrangian submanifolds'' 
hep-th/0003242} 
\lref\ms{G. Moore and G. Segal, unpublished.}
\lref\mmms{M. Mari\~no, R. Minasian, G. Moore and A. Strominger, 
``Nonlinear instantons from supersymmetric $p$-branes,'' hep-th/9911206, 
JHEP {\bf 0001} (2000) 005.}  
\lref\thomas{R.P. Thomas, ``Derived categories for the 
working mathematician,'' math.AG/0001045.}
\lref\lerche{W. Lerche, ``On a boundary CFT description of nonperturbative 
${\cal N}=2$ Yang-Mills theory,'' hep-th/0006100.} 
\lref\dcy{I. Brunner, M.R. Douglas, A. Lawrence and C. R\"omeslberger, 
``D-branes on the quintic,'' hep-th/9906200. D.-E. Diaconescu and 
C. R\"omeslberger, ``D-branes and bundles on elliptic fibrations,'' 
hep-th/9910172. P. Kaste, W. Lerche, C. A. L\"utken and 
J. Walcher, ``D-branes on K3 fibrations,'' hep-th/9912147. E. Scheidegger, 
``D-branes on some one and two-parameter Calabi-Yau hypersurfaces,'' 
hep-th/9912188. M. Naka and M. Nozaki, ``Boundary states in Gepner models,'' 
hep-th/0001037, JHEP {\bf 0005} (2000) 027. I. Brunner and V. Schomerus, 
``D-branes at singular curves of Calabi-Yau 
manifolds,'' hep-th/0001132, JHEP {\bf 0004} (2000) 020. 
F. Denef, ``Supergravity flows and D-brane stability,'' hep-th/0005049.}
\lref\diago{D.-E. Diaconescu and J. Gomis, ``Fractional branes and 
boundary states in orbifold theories,'' hep-th/9906242.} 

\Title{\vbox{\baselineskip12pt
\hbox{RUNHETC-2000-26}
\hbox{hep-th/0006189}
}}
{\vbox{\centerline{BPS states and algebras from quivers}}}

\centerline{Bartomeu Fiol and Marcos Mari\~no}

\bigskip
\medskip

{\vbox{\centerline{ \sl New High Energy Theory Center}
 \centerline{\sl Rutgers University}
\vskip2pt
\centerline{\sl Piscataway, NJ 08855, USA }}
\centerline{ \it fiol, marcosm@physics.rutgers.edu }}

\bigskip
\bigskip
\noindent

We discuss several aspects of D-brane moduli spaces and BPS spectra 
near orbifold points. We give a procedure to determine the decay products 
on a line of marginal stability, and we define the algebra of BPS states in 
terms of quivers. These issues are illustrated in detail in the case of 
type IIA theory on $\IC^2/\IZ _N$. We also show that many of these results 
can be extended to arbitrary points in the compactification 
moduli space using $\Pi$-stability. 

\noindent

\bigskip

\Date{June, 2000}

\listtoc \writetoc
\newsec {Introduction}

D-branes provide a window to non-perturbative aspects of string theory. For
compactifications with enough supersymmetry, the spectrum is well understood.
On the other hand, obtaining the full non-perturbative spectrum of type II 
on Calabi-Yau varieties is an open challenge. The cohomology of the variety,
or more generally K-theory \kpapers, provides a relevant piece of 
information, since it determines the allowed charges of D-branes, or in 
other words, the charge lattice, but finding which sites on that lattice are 
actually occupied and with which degeneracy is a harder problem.

Recently, a framework for the study of classical BPS branes at 
arbitrary points in the compactification moduli space of type II theory on 
Calabi-Yau varieties has started to emerge \refs {\pistab,\dfr,\dmike}. The 
natural language in this framework is that of homological algebra 
and category 
theory, D-branes being the objects in a category, and the fermionic zero 
modes of the strings stretching between the branes playing the role of the 
morphisms in the category. Some of the usual notions of this language, as 
sub-objects or extension groups, 
have already found a role in the physics, but there is still a long 
way to go to develop a complete physical intuition about this new language. 
In this sense, orbifold points seem specially helpful: many of the problems 
we are interested in ($e.g.$ existence or not of boundstates with given 
charges, determination of lines of marginal stability) get reduced to 
questions in linear algebra, and at these corners of moduli space the 
abstract concepts of the general setup boil down to simple manipulations 
of matrices. The mathematics relevant for this family of problems goes by 
the name of quiver theory \qpapers. One of the purposes of the present paper 
is to show how the abstract notions of the general framework are easily 
visualized when we work near orbifold points.

Now that the tools to obtain the (classical) spectrum of BPS branes on 
Calabi-Yau compactifications are starting to be unraveled, it seems 
appropriate to rethink what are the likely lessons to be learned from the 
knowledge of this spectrum. A direction we feel is begging for further 
exploration is based on the following fact: BPS states form an 
algebra \hm\bpsproc. 
What is the significance of this algebra? An analogy that we have in mind are
the chiral rings of ${\cal N}=2$ conformal theories \chiral. In those 
theories one focuses 
on a particular set of states, the chiral primaries, that 
form a closed algebraic structure, and encode a good deal of information 
about the full theory. One might wonder to what extent the algebra of BPS 
states can play a similar role \foot {In \bpsproc, it is 
shown that the algebra of BPS states of a two-dimensional ${\cal N}=2$ 
model is deeply related to  
the chiral ring.}.

Our final goal is to find the quantum spectrum of D-branes, and the classical
spectrum is just a first step in that direction. For true bound states, 
finding the degeneracy involves the quantization of the collective 
coordinates, and the BPS spectrum turns out to be 
given by the cohomology of the 
classical D-brane moduli space\foot{For boundstates at threshold one has 
to be more careful and restrict for example to the small diagonal \hm. A 
full quantum mechanical treatment would involve a study along the lines of 
\sethi.}. So one issue we must address is how to obtain D-brane moduli 
spaces at arbitrary points in compactification moduli space. In practice, as 
we will discuss extensively, constructing moduli spaces involves a notion of 
stability for the configurations we are considering. In this paper we 
discuss in detail how to construct D-brane moduli spaces starting from the 
stability conditions. A key point here is the notion of S-equivalence, which 
gives the right framework to treat objects that are not stable, 
but only semistable. We also 
discuss these issues for the $\Pi$-stability condition, which 
was proposed in \pistab\ as the criterion for stability valid at 
arbitrary points in moduli space. In particular, we show that 
S-equivalence can be also introduced in the context of 
$\Pi$-stability. 

Another aspect of $\Pi$-stability is that it predicts the lines of marginal
stability in the compactification moduli space. We will extend
this proposal by giving a procedure to obtain also the ``decay products'' 
whenever the BPS spectrum jumps by crossing a line of marginal stability. 

After this study of the moduli spaces of D-branes, we define the algebra of 
BPS states near an orbifold point using quiver theory and representations 
of quivers. To do this 
we adapt the correspondence conjecture of Harvey and Moore \hm\ to the 
orbifold point. In this regime, the correspondence conjecture 
involves only linear algebra and the computations become considerably 
easier than in the large volume limit. 

All these notions are illustrated in detail in the case of type IIA 
theory on $\IC^2/\IZ_N$. This is a well-known example which has been 
studied from many points of view, so it is a very good testing ground 
for the above ideas. First we analyze the BPS spectrum of D2-D0 bound 
states by looking at stable representations of the quiver, and we reproduce 
the physical expectations coming from M-theory \md. We also show that, 
although, as expected, there are no jumps in the spectrum, there is a rich 
structure of lines of marginal stability. Next we consider the BPS algebra 
associated to these states, formulated in terms of representations of 
quivers. Because of heterotic/type II duality and the 
results of \hm, this algebra 
should be the subalgebra 
of positive roots of the affine $SU(N)$. We show that this is the 
case in some nontrivial examples by direct computation from the 
correspondence conjecture.  

The organization of the paper is as follows. In section 2 we review some
aspects of the construction of D-brane moduli spaces, emphasizing the notion
of stability. We extend this well-known construction to arbitrary points in
moduli space, using the recent proposal of $\Pi$-stability \pistab. As a 
welcome spin-off, we extend this proposal, providing now a way to obtain the 
``decay products'' along a line of discontinuity of the spectrum. In section
3 we revisit the definition of algebras of BPS states \hm, reformulating it
near orbifold points in quiver language. We also comment on the possible 
non-renormalization of BPS algebras with 16 supercharges. In section 
4 we illustrate many of these ideas in a simple example: we deduce the 
spectrum and algebra of BPS states of type IIA on $\IC^2/\IZ_N$, and 
reproduce what is known on physical grounds. Finally, we state our 
conclusions and prospects for future work.

\newsec {D-brane moduli spaces}

\subsec{The general picture}

Since our discussion will be happily jumping from one point to another in 
compactification moduli space, before descending into the details, we would 
like to provide an eagle's view of the general landscape.

We want to describe the spectrum of BPS D-branes for type II string
theory compactified on a Calabi-Yau variety (see \dtopics\ for background; 
some recent references are \dcy). A basic feature is that we have two types of D-branes 
\refs {\bbs,\ooguri} : A-type, wrapping special
Lagrangian submanifolds in the large volume, and B-type, wrapping holomorphic
cycles. Since much more is known about holomorphic cycles, we focus on B-type
branes, although one expects that eventually, by mirror symmetry, all the
results will have a translation to the A-type branes. Some results for A-type
branes have appeared in \shamit .

Locally, the compactification moduli space splits into complex structure 
moduli space times K\"ahler moduli space. We will fix the complex structure,
and consider the spectrum of BPS branes at different points in K\"ahler
moduli space. In particular, two kinds of points are under specially good 
control: the large volume limit and orbifold points. In the large volume, 
D-branes carry a holomorphic vector bundle, so its classification is related
to that of holomorphic bundles, with extra constraints. Near orbifold 
points, D-branes are described by representations of quivers \dm. In some 
cases one can construct a map between these two classes of objects, 
and this was first shown in \diago. 

Beyond knowing how to describe D-branes themselves, we need to introduce a 
notion of {\it stability}. This idea of stability serves a double purpose. 
First, it allows us to discuss the lines of marginal stability, where the 
BPS spectrum can jump. Second, it provides a practical way to construct 
D-brane moduli spaces, bypassing the need of solving all the equations that 
define the vacua. In the large volume, this notion of stability is known
as $\mu$-stability, and it depends on the Chern classes of the bundle. Near
orbifold points the relevant notion is $\theta$-stability.

Once we leave the safe havens of the large volume or the orbifold points, 
we are in uncharted waters. It is time to admit that we don't know what kind
of beasts are D-branes in the middle of moduli space! However, the 
understanding of the two points in moduli space mentioned before gives some 
clues. The minimal answer seems to be that D-branes are objects in a category 
defined by a holomorphic constraint \refs {\dsharpe,\pistab,\dmike}\foot {
The role of categories for heterotic string compactifications has been
discussed in \daspin. Related ideas are developed in \ms.}. Recall that 
a category is just a set of objects and 
maps between them (see \gm\ for a very clear introduction to category theory).
 D-branes are to be thought as the objects of a category, and the fermionic 
zero modes of the 
strings stretching between them as the maps of the category. This might 
sound very vague, and certainly it is as it stands, but as we will see, 
further physical input will help us to narrow down the kind of category we 
should be considering.

Furthermore, we need to extend the notion of stability to the whole of 
K\"ahler moduli space. In this sense, a proposal was presented in \pistab, 
where it was dubbed $\Pi$-stability, since it involves the periods $\Pi (u)$ 
of the Calabi-Yau variety. It was proven in \pistab\ that $\Pi$-stability 
reduces to $\mu$-stability and $\theta$-stability in the corresponding 
limits. The general picture is displayed in figure 1.

\ifig\moduli{D-branes and stability at generic points in 
K\"ahler moduli space and in particular limits.}
{\epsfxsize5.0in\epsfbox{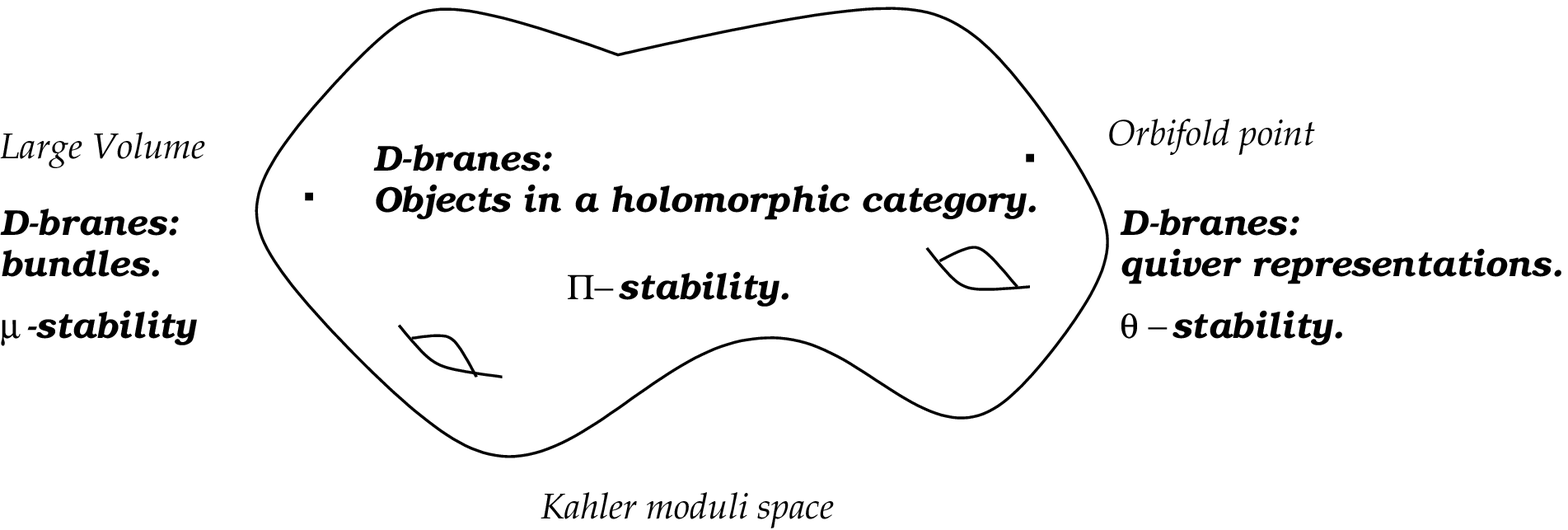}}

\subsec {D-branes and quivers.} 

As shown in \dm, the description of D-branes near orbifold points involves 
supersymmetric gauge theories constructed from quivers. A quiver is simply a  
graph with a set of vertices or nodes $V$, and a set of arrows $A$ 
going from node $i$ to 
node $j$. For each arrow $a$ we denote by $ia$ its initial node and by
$ta$ its terminal node. Given a quiver, one can obtain a gauge theory 
by considering a {\it representation} of the quiver. This quiver gauge 
theory encodes the information about the classical 
boundstate spectrum and the lines of marginal stability near orbifold points.
In \refs {\pistab,\dfr} it was pointed out that quiver theory allows to 
extract that information. Let's recall then the basic definitions.

A representation of a quiver is given by a (complex) vector space for 
each node and a matrix for each arrow: $\{U_v, v=1,\dots ,V;\;\phi _a: 
U_{ia}\rightarrow U_{ta}\}$. The dimension vector of a representation is a 
vector with $V$ components given by the dimensions of the vector spaces: 
$n =(\hbox {dim}\, U_1,\dots ,\hbox {dim}\, U_V)$. Notice that a 
representation of the quiver corresponds to a configuration of an  
${\cal N}=1$ supersymmetric gauge theory with gauge group $G=\prod_{v \in V} 
U(n_v)$, and with chiral 
multiplets $x_a$ associated to the different arrows between the nodes. These 
multiplets are in the bifundamental representation 
$(n_{ta},\overline n_{ia})$.  

Since the configuration space of the ${\cal N}=1$ gauge theory 
is given, roughly speaking, by the space of representations with a 
fixed dimension, it is important to understand this space in some detail. We 
will then give some mathematical definitions which will be very useful.
  
We will say a representation is {\it indecomposable} if 
it can not be written as direct sum of two representations. 
It can happen that a representation is indecomposable, but ``small 
perturbations'' of it are not. When in a neighborhood of an indecomposable
representation all representations are indecomposable, we call it a {\it 
stably indecomposable} representation, or {\it Schur} representation. One 
can prove that a representation $R$ is Schur if and only if 
${\rm End} \, R= \IC$. The importance of Schur representations stems from 
this property, since it implies that D-brane boundstates are given by Schur 
representations \dfr. The dimension vector of a Schur representation is 
called a {\it Schur root}. 

Given two representations $R$ and $S$, we can define an homomorphism between 
representations by a collection of linear maps, one per node, $\phi (v): R(v)
\rightarrow S(v)$ such that for all arrows, $\phi (R(a))=S(\phi (a))$.
If the homomorphism $\phi: R\rightarrow S$ is injective, we say that $R$ is
a {\it subrepresentation} of $S$. For any $\phi$, ${\rm Im}\;\phi$ is a 
subrepresentation 
of $S$ and ${\rm Ker}\;\phi$ is a subrepresentation of $R$. if the generic 
representation with dimension vector $n$ has a subrepresentation with 
dimension vector $n'$, we say that $n'$ is a {\it subvector} of 
$n$. As we will see, the relevance of the subrepresentations of 
a representation is that as we wander in moduli space, subrepresentations 
are the candidates to destabilize the original representation, causing 
the BPS spectrum to jump \pistab.
  
\subsec {$\mu$-stability and $\theta$-stability}

Both at the orbifold and in the large volume, BPS configurations 
are given by solutions to
two kinds of equations, modulo gauge equivalence by the 
appropriate gauge group $G$. The first one is a 
holomorphic constraint: in the large
volume it implies that we have to restrict to holomorphic bundles, 
$F^{2,0}=0$, and at the orbifold point, one has the F-flatness conditions 
coming from the superpotential, a holomorphic quantity. 

The second kind of equation imposes the BPS condition. Near the orbifold point
this is the D-flatness condition, and in the large volume, the equation is
$F^{1,1}=\zeta \omega$ or some deformation thereof \mmms, where 
$\omega$ is the K\"ahler form. It turns 
out that both these equations can be 
reformulated as a (quasi)-topological condition\foot {The reason that we say 
(quasi)-topological is that it depends explicitly on K\"ahler moduli.}.
These reformulations at very different points in moduli space share many 
common properties, and lead naturally to a generalization of the 
(quasi)-topological condition for arbitrary points in moduli space. 

This reformulation arises as follows. In order to solve the equations, it 
proves useful to solve first the complex one. For example, 
in the large volume limit, this 
corresponds to considering connections that define a holomorphic bundle. 
Second, since the space of solutions to the complex equations 
is invariant under the 
action of the {\it complexified} gauge group $G_{\IC}$, one can talk 
about the $G_{{\scriptstyle \IC}}$-orbits satisfying the complex equations. 
The key result 
is that, under certain conditions, these complex orbits contain a solution of 
the real equation. This condition is called {\it stability}.       
  
In the large volume limit, stability refers to the usual 
Mumford stability condition for holomorphic vector 
bundles (see, for example, \refs{\koba,\potier}, and also \sharpe\ for a 
physical point of view): for a bundle $F$ of rank $r(F)$ on a K\"ahler 
manifold, define its degree as 
\eqn\degree{
{\rm deg}(F)=\int c_1(F)\wedge \omega^{d-1},} 
and its slope $\mu$ as
\eqn\slope{
\mu(F)={ {\rm deg}(F)\over {\rm r} (F)}.}
The bundle $F$ is $\mu$-semistable if and only if for every 
subbundle $F'$ of $F$ we have 
\eqn\subb{\mu(F')\leq\mu (F).}
If $\mu(F')<\mu (F)$ we say that $F$ is $\mu$-stable. Note that $\mu(F)$ 
depends explicitly on K\"ahler moduli through $\omega$. The relevance of 
$\mu$-stability is the following: if the bundle is holomorphic, the 
holomorphic structure is defined by a $G_{\IC}$-orbit of gauge connections 
satisfying $F^{2,0}=0$. If the bundle is stable, then there is one 
representative in this orbit which solves the real equation $F^{1,1}=
\zeta \omega$.

Near the orbifold point, the world-volume of D-branes 
is given by quiver gauge 
theories \dm, and D-brane configurations correspond to representations
of quivers which satisfy the F-flatness and D-flatness conditions. We write 
these conditions schematically as:
\eqn\dfconds{{\partial W(\phi)\over \partial \phi}=0\,\,\,\,\,\,\,\sum [\phi,
\phi ^\dagger]=\theta,}
where $W(\phi)$ is the superpotential depending on the chiral fields $\phi$. 
In the second equation, the right hand side is given by a 
block-diagonal matrix 
whose entries are  $\theta_v \cdot {\bf 1}_{n_v \times n_v}$, 
$v=1, \cdots, V$. In this case, the appropriate notion of stability 
is $\theta$-stability \king. Let $\theta =
(\theta_1, \cdots, \theta_V)$ be a vector whose V components are real 
numbers. A representation of a quiver is called $\theta$-semistable if 
$\sum_v n_v \theta_v=0$, and for every subrepresentation with dimension 
vector $ n'$, one has:
\eqn\subr{\sum_{i}n'_i \theta_i \ge \sum_{i}n_i \theta_i=0.}
If $\sum_{i}n'_i \theta_i>0$ we say that the representation is 
$\theta$-stable. As in the large volume limit, the procedure to 
solve the equations \dfconds\ is to solve first the complex ones, and 
consider complexified orbits of the solutions. The complexified 
gauge group is $G_{\IC}=\prod_v {\rm Gl}(n_v, \IC)/D$, where we quotient 
by the diagonal $D=(\lambda \cdot {\bf 1 }_{n_1\times n_1}, \cdots, 
\lambda \cdot {\bf 1}_{n_V \times n_V})$, $\lambda \in \IC^*$.
 A theorem of King \king\ 
guarantees that, if the representation is $\theta$-stable, 
the $G_{\IC}$-orbit 
will contain a solution to the D-flatness conditions.  

The vector $\theta$ is closely related to the physical 
Fayet-Iliopoulos (FI) terms. For $\IZ_N$-orbifolds, the relation is given 
by \pistab:
\eqn\valor{
\theta = \zeta -{ \zeta \cdot  n \over  e \cdot n} e,}
where $e =(1,1, \cdots, 1)$, and $\zeta=(\zeta_1, \cdots, \zeta_N)$ 
are the FI terms \foot{In this paper, we will always choose $\zeta_{\IC}=0$, 
so $\zeta=\zeta_{\IR}$.}. Notice 
that $\sum_v \zeta_v=0$, therefore 
$\theta \cdot n=0$. The condition \valor\ is obtained after taking into 
account that the D-flatness conditions of \dfconds\ describe in fact 
quasi-supersymmetric vacua with a constant, nonzero value for the 
vacuum energy \ddg. Requiring this energy to be minimal in the 
sector specified by the dimension vector $n$ gives \valor. When written 
in terms of the Fayet-Iliopoulos terms, $\theta$-semistability becomes:
\eqn\thetass{
 { \zeta \cdot  n' \over  e \cdot n'} \ge { \zeta \cdot  n \over  e \cdot n},}
for any subrepresentation of dimension $n'$. Notice the 
formal similarities between $\mu$-stability and $\theta$-stability: in both 
cases we have an object (a bundle or a quiver representation), and to satisfy 
stability it has to obey an inequality (eqs. \subb\ or \subr) against the
list of all its subobjects (subbundles or subrepresentations). In fact, 
$e\cdot n$ plays the role of ${\rm r}(F)$, while $-\zeta \cdot n$ plays the 
role of ${\rm deg}(F)$. 

\subsec{$\Pi$-stability}

In view of the previous considerations, we are led to 
ask if there is a generalization of these stability criterions that is valid 
everywhere in moduli space, and reduces to $\mu$ or $\theta$ stability in
the corresponding limits. Since we want a criterion exact in $\alpha'$, if
it exists at all, it is natural to suspect that it will involve the periods
$\Pi(u)$, since they can be exactly determined using mirror symmetry.

To start with, inspired by the large volume and orbifold examples, it was
assumed in \pistab\ that D-branes at generic points in moduli space have also
sub D-branes (the generalization of sub-bundles and subrepresentations), so
we require that there is a notion of subobject of an object in our category.
The technical definition is that $E'$ is subobject of $E$ if there exists
an injective homomorphism from $E'$ to $E$. With this
assumption, the proposal of $\Pi$-stability is the following \pistab: at a 
point $u$ in K\"ahler moduli space, define the {\it grading} of a brane $E$ 
with central charge $Z(E,u)$ \foot {As explained in detail in \pistab\ , it 
is necessary to extend slightly the usual notion of central charge, by 
keeping track of the integer part of the phase as we wander in moduli space
$$Z(E,u)=\Pi(u)\cdot Q(E)\ e^{2\pi in(E,u)}$$.} by
\eqn\grade{
\varphi (E,u)={1\over \pi} {\rm Im}\ \log\ Z(E,u)}
We say that $E$ is $\Pi$-semistable at $u$ if and only if for all the 
sub-branes $E'$ of $E$ at $u$ we have 
\eqn\desig{
\varphi (E)\geq \varphi (E').}
If we have strict inequality, we say that $E$ is $\Pi$-stable. It was shown 
in \pistab\  that this definition reduces to $\mu$-stability in the large 
volume and to $\theta$-stability at orbifold points. This is displayed in 
figure 1 above.

Recall that we started with two sets of equations at very different points
in moduli space. Some of these equations admitted a translation into a
stability condition, and it was in this guise that a generalization was 
proposed in \pistab. This raises the interesting question of what is the 
equation that corresponds to $\Pi$-stability, in the same sense 
as the D-term equation gives rise to $\theta$-stability at orbifold points 
or $F^{1,1}=\zeta \omega$ translates into $\mu$-stability near the large 
volume. 

\subsec {Moduli spaces and S-equivalence} 

In the previous
subsection we have reviewed how the D-brane moduli spaces arise as solutions
to classical equations with gauge invariance. In both cases, the usual way
to construct the moduli spaces of solutions is by exploiting the 
correspondence between the solutions and a suitable category of stable 
objects ($\mu$-stable bundles in the large volume, $\theta$-stable
representations at the orbifold point). Let's now 
discuss the orbits which are not stable, but only semistable. 

The situation for semistable orbits is more 
complicated, since some of them do not contain any solution to the 
classical equations, and some of them do. This means that some semistable 
representations are not really relevant to the solution of the moduli 
equations. On the other hand, it is highly desirable to describe this moduli 
space in terms of representations obeying (semi)stability conditions. 
The solution to this problem is to introduce an equivalence 
relation in the set of semistable 
orbits in such a way that, for each equivalence class, there is one and 
only one solution to the real equation. This relation is called 
{\it S-equivalence}. Among the representations which are in the 
S-equivalence class, there is one which is the actual solution 
to the moduli equations, and it is called the graded representation. 
As we will see, S-equivalence also provides a recipe to obtain this graded
representation. It is important to stress that S-equivalence, at least in 
this situation, is an artifact of our description: we have chosen 
to describe the moduli space of solutions to our equations in terms 
of holomorphic objects, but in the semistable case some of 
these objects are spurious. S-equivalence is the proper way to 
get rid of the extra objects, by identifying them to the honest solution 
to the original equations. We will    
explain here in some detail the definition of S-equivalence in the 
case of representations of quivers \king. A discussion of S-equivalence in 
the context of  
heterotic compactifications can be found in \refs {\daspin, \sharpe}.

To give a precise definition of S-equivalence, we have to make some 
technical definitions. First of all, we define 
the {\it Jordan-H\"older filtration} of a $\theta$-
semistable 
representation $R$ as an increasing filtration of $\theta$-semistable 
representations $R_i$:
\eqn\filtr{
0=R_0 \subset R_1 \subset R_2 \subset \cdots \subset R_n=R,} 
satisfying two conditions (see for example \nakapaper): first, 
$\theta \cdot ({\rm dim}\,R_i)=0$, where $({\rm dim}\, R _i)$ is the 
dimension vector of $R_i$; second, the quotients $M_i=R_i/R_{i-1}$ are
 $\theta$-stable (in particular, $M_1=R_1$ is a stable subobject of 
$R$). One can prove that any 
semistable representation has a Jordan-H\"older filtration. Notice that, 
if $R$ is $\theta$-stable, then one can take as a 
Jordan-H\"older filtration just $0\subset R$. We now define the 
{\it graded representation} ${\rm gr}(R)$ as:
\eqn\gr{ 
{\rm gr}(R)=\oplus_i  M_i.}
The main point is that, although the Jordan-H\"older filtration of $R$ 
is not necessarily unique, the graded representation of $R$ is unique 
(up to isomorphism). If $R$ is $\theta$-stable, its graded representation 
is $R$ itself. We can now define $S$-equivalence. Let $R$, ${\tilde R}$ be 
$\theta$-semistable representations. We say that $R$ and ${\tilde R}$ are 
S-equivalent (and we write $R \sim_S \tilde R$) if ${\rm gr}(R)$ and 
${\rm gr}(\tilde R)$ are isomorphic representations. There are two 
immediate consequences of this: first of all, two strictly stable 
representations are S-equivalent if and only if they are isomorphic. 
Second, a representation $R$ is always S-equivalent to 
${\rm gr}(R)$. 

Notice that there are representations which are S-equivalent but 
are not in the same $G_{\IC}$-orbit. The $G_{\IC}$-orbit which 
actually contains the solution to the equations is the one which is 
a {\it direct sum} of $\theta$-stable representations \koba. This means that, 
as we have stressed above, given 
a $\theta$-semistable representation $R$, the solution to our 
equations will be in the $G_{\IC}$-orbit of the graded 
representation ${\rm gr}(R)$. We will 
see that the graded representation associated to a representation 
has in fact a very clear physical meaning: it corresponds to the 
``decay products'' or ``primary constituents'' \rusos\ of the BPS state 
represented by $R$ when we are on a line of marginal stability.  

In conclusion, the D-brane moduli space we are looking for can be realized 
as the space of S-equivalence classes of semistable $G_{\IC}$-orbits which 
solve the complex equations. In the orbifold limit, the space of 
$\theta$-semistable representations 
with dimension vector $n$ will be denoted as ${\cal M}_\theta (n)$. Since, 
for a given $n$, $\theta$ is related to $\zeta$ by \valor, we will 
also denote this space by ${\cal M}_\zeta (n)$. 

Although the description of the space of solutions in terms of stable objects
sounds rather formal, it is of practical use: in order to find the 
moduli space of solutions, instead of solving the real and the complex 
equations modulo usual gauge invariance, one can just solve the 
complex equation modulo complexified gauge transformations, and look for 
semistable objects up to 
S-equivalence. We will give examples of this procedure later on. 
 
\subsec{S-equivalence and $\Pi$-stability}

Since our goal is to describe D-brane moduli spaces everywhere in 
K\"ahler moduli space, we should try to make sense of S-equivalence 
for the whole of moduli space. As we 
have explained, S-equivalence requires a notion of stability, and 
$\Pi$-stability is our only current candidate, so we will try to extend the
notion of S-equivalence everywhere in moduli space using $\Pi$-stability.

In doing so, we assume explicitly, following \pistab, that the category
of D-branes is abelian \foot {We would like to make a remark concerning our 
assumption of the category of D-branes being abelian \dmike . There is a
number of papers in the physics literature advocating that derived
categories are the right framework to think about D-branes \refs {\dsharpe, 
\daspin}; in fact, this idea
predates the modern understanding of D-branes, and originates in the work
of Kontsevich on mirror symmetry \mirror. Very roughly, if we think
of D-branes as objects in a category, the derived category would correspond
to identifying D-brane configurations that differ by brane-antibrane pairs. 
The bad news is that in general derived categories are not abelian (see, for 
example, \thomas.)
Therefore, at present, the physical input that we are using to try to pin 
down the precise category that describes D-branes, seems to translate into 
not obviously compatible mathematical requirements.}. First we will present 
the notion of Jordan-H\"older 
filtration for an abelian category. Then we will try to argue that at
arbitrary points in moduli space, the set of $\Pi$-semistable D-branes 
with fixed grading $\varphi$ forms an abelian subcategory of the category of 
all possible D-branes, so the previous generic definition applies to them. The
presence of the integral part of the phase in the grading (see footnote 6)
will prevent us from giving a complete argument, but we will present a proof 
in the case of small difference of phases.

Let ${\cal A}$ be an abelian category. We say that an object $E\in {\cal A}$
is {\it simple} if it does not have any proper subobject. An increasing 
sequence of subobjects $0 \subset 
E_1\subset E_2\subset \dots E_n=E$ is called a 
Jordan-H\"older series of $E$ if $E_1$ and the quotients $M_i=E_i/E_{i-1}$ 
are all simple. Given a Jordan-H\"older series we define
$$
\hbox {gr}(E)=\oplus _i M_i,$$
which is the natural generalization of \gr\ to any abelian category. 

An abelian category is {\it artinian} if every decreasing sequence is finite.
After all these definitions, we are ready to present the theorem that we 
will rely upon

{\bf Theorem} \sesha . Let ${\cal A}$ be an abelian category. If an object
$M$ in ${\cal A}$ has a Jordan-H\"older series, then ${\rm gr} M$ is unique
up to isomorphism. Furthermore, if ${\cal A}$ is artinian, then every object 
in ${\cal A}$ has a Jordan-H\"older series.

Before we discuss the generic case, let us recall the situation at the
orbifold point. There King \king\ argues that the category of all $\theta$-
semistable representations for a fixed $\theta$ is abelian, artinian and
noetherian. Furthermore, the simple objects in this category are the
$\theta$-stable objects: consider a $\theta$-stable object; any proper
subobject should satisfy $\theta \cdot n' >0$, but by definition all the
objects in that category satisfy $\theta \cdot n'=0$.

In the large volume, there is a difference: $\mu$-stable bundles can have
$\mu$-stable sub-bundles, the only requirement being that the slope is smaller
$\mu(E')<\mu(E)$. So the category to look at is that of $\mu $-semistable
bundles with fixed slope $\mu$ \potier. This is also the generic category
to consider at arbitrary points in moduli space: the category 
${\cal A}_\varphi (u)$ of $\Pi$-semistable objects with a fixed grading 
$\varphi$. By the same argument as before, the simple objects in this 
category are the $\Pi$-stable objects. According to the previous theorem, we 
should argue that ${\cal A}_\varphi (u)$ is artinian to prove the existence
of a Jordan-H\"older series, and abelian to prove that its graded sum is
unique. In the following we will argue that it is abelian. We don't have
a rigorous argument showing that it is artinian, but on physical grounds 
this seems quite plausible: in the next subsection, we will argue that the 
graded sum gives the decay products on a line of marginal stability. If our 
proposal is valid, $A_\varphi(u)$ not being artinian would imply that some
objects would decay into an infinite number of subobjects on the line of
marginal stability.

To prove that this subcategory is actually abelian, we can apply almost 
verbatim the proof in \potier\ for bundles in the large volume. There is a
small complication, however; when we have an exact sequence of bundles
\eqn\seqfirst{
0\rightarrow E\rightarrow G\rightarrow F\rightarrow 0,}
the following relation among their slopes follows: 
\eqn\slopes{
\mu (G)=
{\hbox {r}(E)\over \hbox {r}(G)}\mu (E)+{\hbox {r}(F)\over \hbox {r}(G)}
\mu (F).} 

There is no such relation for the gradings $\varphi (E), \varphi (G), 
\varphi (F)$. Although this relation among slopes is used repeatedly in 
\potier, we don't need that much. When the difference of phases is small, 
the gradings satisfy a convexity condition
\eqn\conv{
\varphi (G)=x\varphi (E)+(1-x)\varphi (F)\;\;\;\hbox {  for some } 
0\leq x \leq 1}
(the relation for the slopes is a particular case of this), and this turns out
to be enough for the proof to go through\foot {We are grateful to M.R. Douglas
for this suggestion.}. The detailed proof is presented in the appendix. 

\subsec{Bound states and lines of marginal stability: a proposal 
for decays}

According to the proposal of \pistab\ and \dfr, BPS bound states at the 
orbifold point are described by $\theta$-stable representations, where the 
value of $\theta$ is related to the physical FI parameters by \valor. 
Since $\theta$-stable representations are Schur, this is in agreement with 
the idea that bound states of D-branes leave a single $U(1)$ unbroken 
\foot{Notice that a representation 
can be Schur and nevertheless be unstable for some values of the 
parameters, and therefore does not correspond to a (quasi)-supersymmetric 
vacuum.}. 
In general, however, there will be a whole moduli space of 
$\theta$-semistable representations ${\cal M}_{\zeta}(n)$. In the 
semiclassical approximation, one has to do supersymmetric quantum 
mechanics on this moduli space, and the spectrum of BPS states will be 
given by the $L^2$-cohomology of the moduli space, as in \hm:
\eqn\hilbert{
{\cal H}_{\rm BPS}^{n,\zeta} =H^{*}_{L^2}({\cal M}_{\zeta}(n)).}
We have recorded the dependence of this Hilbert space on the 
background FI parameters. We should say that the above equation is 
still very incomplete. For example, as noted in \hm, to understand 
in detail the multiplicity structure of the BPS states it is necessary 
to consider representations of the supertranslation algebra. However, 
\hilbert\ will be enough for our purposes in this paper. 

An important feature of the spectrum of BPS states for a theory with 8 
supercharges is that it can present discontinuities as we move in moduli 
space. This was first observed in two dimensions \ceco, and plays an 
important role in the celebrated solution by Seiberg and Witten of 
${\cal N}=2$ SU(2) SYM \sw . It is well known that a necessary condition for 
the existence of lines of discontinuity in the spectrum is given by
\eqn\marg{
{\rm Im} {Z_1 (u) \over Z_2(u)}=0,}
where $Z_{1,2}(u)$ are central charges at the point $u$ in moduli space. The 
question of when the spectrum does actually jump is much harder. 
$\Pi$-stability provides an answer: the spectrum will jump when,
 given a vector
charge $Q$, the cohomology of the moduli space of $\Pi$-stable objects 
with that charge changes when passing through the line of marginal stability. 
This means that the moduli space of stable objects has changed, due to the 
fact that objects that were stable on one side of the line 
of marginal stability become unstable on the other side. Therefore, 
the lines of marginal stability have a very clear 
mathematical counterpart: they correspond to the values of the 
background parameters for which stable objects become 
semistable. At the lines of marginal stability, the best we can have is a 
BPS state at threshold. To know 
if this is the case we should address the full quantum-mechanical problem, 
as in \sethi. In this paper we will only use semiclassical considerations, 
so we are not going to tackle this issue. 

Let us now focus on the situation at the orbifold point, and 
consider for simplicity the case in which ${\cal M}_{\theta}(n)$ 
is zero-dimensional and consists of a  
$\theta$-stable representation $R$. In general, in the parameter space 
of $\theta$'s there are 
lines of marginal stability where $R$ becomes 
just $\theta$-semistable. In this case, there will be a nontrivial 
S-equivalence class of representations, including (at least) 
$R$ itself and its graded representation: 
$[R]_S=\{ R, {\rm gr}(R), \cdots \}$. As discussed in \sharpe\ for the large
volume, as we move away from the line of marginal stability, and depending 
on the region we move to, some of the representations in $[R]_S$ 
will become stable, while the others will become unstable. There 
are two possible outcomes of this process \foot{While this paper was being 
typed, the paper \rusos\ appeared, where these two possible outcomes 
were also discussed from a dynamical point of view.}: 

1) It can happen that, as we move from the line, there is a 
$\theta$-stable representation with the same charge. This means that  
the corresponding BPS state exists on both sides of the line of marginal 
stability. This will be in 
fact the case for the examples studied in this paper.   

2) If there is no $\theta$-stable representation with the 
given charge, we clearly have a 
jump in the BPS spectrum. This is interpreted sometimes (see for 
example \bilal) by saying that the BPS state has decayed through 
the line. In this case, we propose that the decay products are 
given precisely 
by the graded representation:
\eqn\decay{
R \longrightarrow {\rm gr}(R)=\oplus_i M_i}
The jump in the spectrum can be interpreted by saying, as in \pistab, that 
the object $R$ has been destabilized by a subobject $R_1$ (notice that $R_1$ 
comes from the Jordan-H\"older filtration \filtr\ and therefore it is 
always a subobject of $R$). Moreover, the graded representation
 gives the full content of the decay products, and not only a 
destabilizing subobject. This proposal is very natural, since the 
graded representation gives the decomposition of the representation 
into the ``minimal'' objects, {\it i.e.}, into the stable pieces 
that constitute the semistable object. This decomposition 
is in fact consistent with the central charge criterion, 
and as we will see later in some examples, \decay\ implies that  
\eqn\cendes{
||Z(R)||=\sum_i ||Z(M_i)|| .}
The physical mechanism behind the ``decay'' process has been explained 
more precisely in \rusos. 
In their language, the products of the decay are in fact 
``primary constituents'' of the BPS state which are no longer 
bound. Our proposal says which of the possible primary constituents 
of the object are actually becoming unbound at the line of marginal 
stability. 
    
The same discussion applies to the 
large volume limit, where the BPS states are associated to bundles. 
Again, semistable bundles on the line of marginal stability are 
S-equivalent to graded bundles. Since we also have 
Jordan-H\"older filtrations and graded objects in the 
context of $\Pi$-stability, the proposal extends in fact to the whole 
of the moduli space. 

\ifig\jordanfig{Before arriving to the line of marginal stability, $R$ is
stable. On the line it becomes semistable, and S-equivalent to $M_1\oplus 
M_2$.} {\epsfxsize1.5in\epsfbox{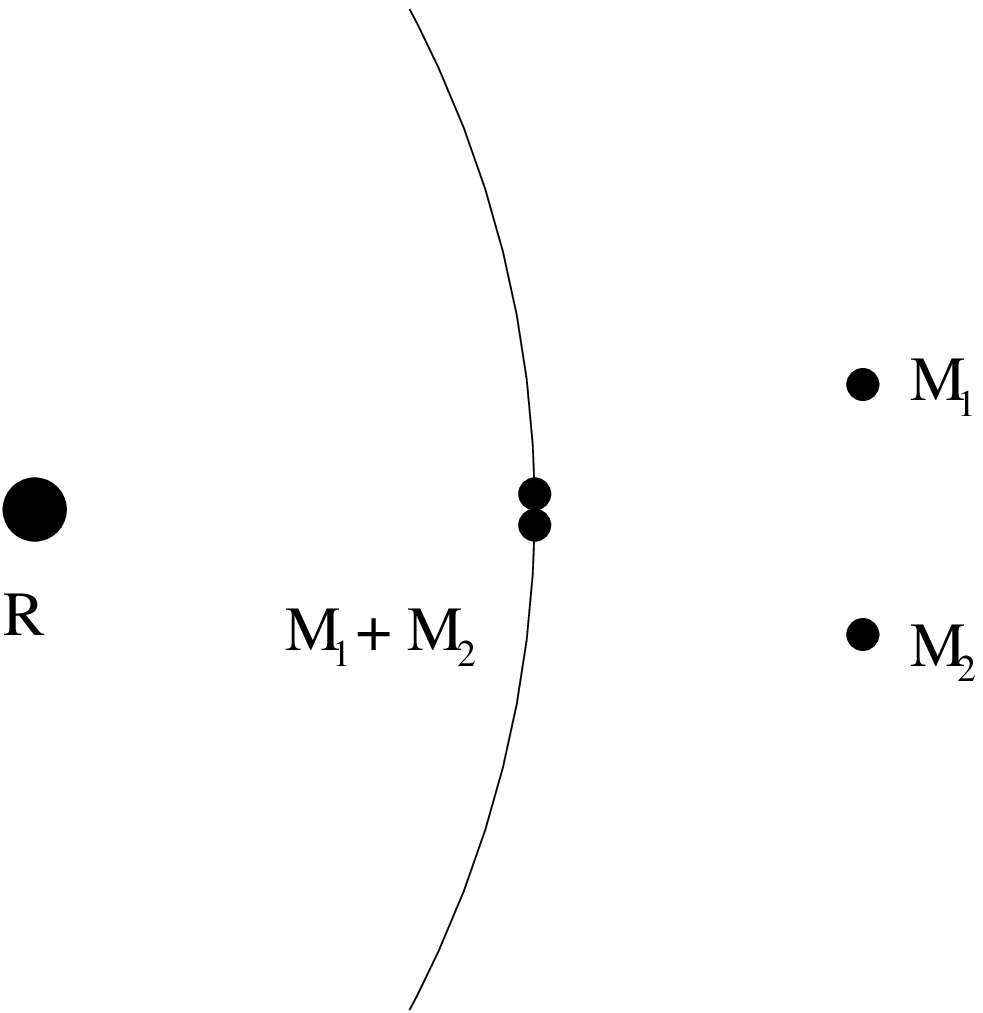}}

In the discussion above we have considered for simplicity moduli 
spaces involving one single representation. The argument extends 
easily to the case in which one has a finite number of representations. 
When the moduli space has positive dimension and the BPS states 
are given by the $L^2$-cohomology, the situation is much more 
subtle. Probably, there are extra quantum numbers associated to some 
group action in the cohomology (like in \refs {\phases,\gv}) that have to be 
taken into account more carefully.    
      
\newsec {BPS algebras from quivers}

\subsec{Definition of the BPS algebra}

It was shown in \hm\ that BPS states form an algebra which captures 
some important nonperturbative aspects. In the large volume limit of 
type IIA compactifications, the BPS states associated to wrapped D-branes 
are described in terms 
of bundles (or, more generally, of coherent sheaves). The proposal of \hm\  
for the computation of the BPS algebra in this regime is based 
on the so called ``correspondence conjecture.'' 
The correspondence conjecture is in general difficult to test, 
since it involves moduli spaces associated to short exact sequences 
of bundles. However, it turns out that this conjecture can be 
adapted readily to the orbifold point, and more generally, to any 
abelian category with a notion of stability. We will then adapt 
the construction of \hm\ to define the algebra of BPS states in terms 
of representations of quivers. The big advantage of the 
orbifold point is that computations can be 
done using linear algebra, and very little geometry. 

Consider a quiver, and two dimension vectors $n_1$ and $n_2$, as well 
as their sum 
$n_3=n_1+n_2$. Consider the moduli spaces of semistable 
representations of this quiver for the 
different dimensions, ${\cal M}_{\zeta}(n_i)$. We will 
assume that $\zeta$ is chosen in such a 
way that all the points in these spaces are in fact isomorphism classes 
of stable representations, and that they are nonempty. We now define the 
correspondence variety ${\cal C}(n_1,n_2;n_3)$ as follows:
\eqn\corr{
 {\cal C}(n_1,n_2;n_3)=\{ (R_1, R_3, R_2) 
\in \prod {\cal M}_{\zeta}(n_i) 
: 0 \rightarrow R_1 \rightarrow R_3 \rightarrow R_2 \rightarrow 0\}.}
The correspondence variety is then given 
by the set of isomorphism classes of triples that fit into 
an exact sequence.
 
{\bf Remarks}: 

1. As it stands, this definition is incomplete, 
since we are considering 
only stable objects. One should also include S-equivalence 
classes of semistable objects, by extending the above definition in 
the obvious way. However, one should check that the definition 
makes sense, 
{\it i.e.} that if $R_{1,2,3}$ fit into a short exact sequence, and $R'_i
\sim_S R_i$, then $R'_{1,2,3}$ fit too. We suspect that this is the 
case, but we do not have a proof. Notice that the same caveat applies 
to the definition in \hm\ in terms of bundles.

2. It is easy to prove that, if $R_3$ is stable, 
then one can have the short exact sequence $0 \rightarrow R_1 \rightarrow 
R_3 \rightarrow R_2 \rightarrow 0$ or $0 \rightarrow R_2 \rightarrow 
R_3 \rightarrow R_1 \rightarrow 0$, but not both. To see this, notice that 
\eqn\sumrule{
(e\cdot n_1) \Bigl\{ {\zeta\cdot n_3 \over e \cdot n_3} - 
{\zeta\cdot n_1 \over e \cdot n_1} \Bigr\} + 
(e\cdot n_2) \Bigl\{ {\zeta\cdot n_3 \over e \cdot n_3} - 
{\zeta\cdot n_2 \over e \cdot n_2} \Bigr\}=0.} 
This is the orbifold analog of the equality \slope\ for the slopes. 
Therefore, if one has the short exact sequence 
$0 \rightarrow R_1 \rightarrow R_3 \rightarrow R_2 \rightarrow 0$, and 
$R_3$ is stable, then $R_2$ cannot be a subrepresentation of $R_3$, since 
this would violate stability.   

We are now ready to define the ``correspondence product.'' Recall that 
BPS states are associated to $L^2$-cohomology classes of 
${\cal M}_{\zeta}(n_i)$. Let 
$\omega_i \in H_{L^2}^*({\cal M}_{\zeta}(n_1))$,   
$\eta_j \in H_{L^2}^*({\cal M}_{\zeta}(n_2))$, $\psi_k    
\in H_{L^2}^*({\cal M}_{\zeta}(n_3))$ be a basis of cohomology classes 
in the Hilbert spaces. We define the BPS product as in \hm:
\eqn\bpsprod{
\omega_i \otimes \eta_j =\sum_k {\cal N}_{ij}^k\,  \psi_k,}
where the coefficients ${\cal N}_{ij}^k$ are given by:
\eqn\coeff{
{\cal N}_{ij}^k=\int_{{\cal C}(n_1,n_2;n_3)} \psi_k^* \wedge \omega_i 
\wedge \eta_j.}
In the above integral, the differential forms on the moduli spaces 
${\cal M}_{\zeta}(n_i)$ are pullbacked to the correspondence variety 
in the natural way. The above product defines an algebra structure 
on the space of BPS states, as in \hm. Given this product 
we define a bracket as 
follows:
\eqn\brac{
[\omega_i, \eta_j] = \omega_i \otimes \eta_j - \eta_j \otimes \omega_i = 
\sum_j c^k_{ij}\,  \psi_k,}
where the structure constants are given by: 
\eqn\str{
c_{ij}^k= {\cal N}^k_{ij} - {\cal N}_{ji}^k.}  

{\bf Remarks}:

1. When the correspondence variety has zero dimension (and this 
will be the case in some important examples) the coefficients 
${\cal N}_{ij}^k$ are just integers that count the number of short 
exact sequences. In that case the algebra that we have defined 
is essentially given by the Ringel-Hall algebra 
associated to the quiver diagram (see \refs {\ringel\ringeltwo}\ 
and also \frenkel\ for a recent discussion). 
When the moduli space has positive dimension, the 
algebra of BPS states, as defined above, should be 
closely related to the algebras defined in this 
context by Nakajima \nakalg.  

2. Our definition is a little bit different from the one 
given in \hm\ in terms of bundles, since we make a 
distinction between the product \bpsprod\ and the 
bracket \brac. However, due to our remark 2 above, in many cases 
$N_{ij}^k$ and ${\cal N}_{ji}^k$ cannot 
be both nonzero, so the product \bpsprod\ and the bracket agree up 
to a sign (a similar situation arises in Ringel-Hall algebras, see 
\ringeltwo).

3. The BPS algebra has a direct physical interpretation: if $c_{ij}^k \not= 
0$, then the BPS states associated to $\omega_i$ and $\eta_j$ can form 
the boundstate $\psi_k$. Therefore, the BPS algebra carries 
information about the structure of bound states of the theory. 
Notice that, if $c_{ij}^k \not=0$, then one has in particular that 
${\rm Ext}(R_1,R_2)\oplus {\rm Ext}(R_2,R_1) \not=0$, which is a necessary 
condition to have a boundstate pointed out in \pistab. However, the 
criterion based on the BPS algebra structure is more precise, since one 
can have a nontrivial extension without having a boundstate (for example, if 
the extension is unstable).

4. The values of ${\cal N}_{ij}^k$ and $c_{ij}^k$ depend 
in general on the values of the FI parameters $\zeta$. In the examples 
the we will consider later on, this dependence is rather mild, and 
will be given by a change of sign in some of the structure constants 
when we pass through the lines of marginal stability. 

5. The algebra is computed in the approximation where BPS states are 
identified with cohomology classes in some moduli space. In this 
approximation, the computation is purely cohomological. In particular, 
there is no trace of the string coupling constant $g_s^{II}$. 
More generally, one can ask what is the dependence of the BPS algebras on 
the coupling constant. We would like to add some comments on this issue.

\subsec {Renormalization of the algebra of BPS states with 16 supercharges}

In \hm, Harvey and Moore raise the question whether BPS algebras with 
16 supercharges don't get renormalized and present some evidence for
BPS states annihilated by the same supersymmetry generator, in
the case of type IIA on K3/ Het on $T^4$. This is easier to analyze
from the heterotic side.

In toroidal compactifications of the heterotic string, the perturbative BPS
states are the Dabholkar-Harvey states \dh . The rightmoving 
(supersymmetric) sector is in the ground state, whereas the leftmoving 
(bosonic) sector is arbitrary, only constrained by level matching. Because
the rightmoving sector is identical to those of massless states, it is 
conceivable that the world-sheet non-renormalization arguments given
for massless scattering amplitudes extend for BPS states annihilated by
the same supercharges, or maybe even arbitrary BPS states \dh. 
Indeed, note that the 2-point function of these states is exact. 
More to the point, since their algebra is extracted from the 3-point 
functions \hm, if the world-sheet 
non-renormalization theorems generalize, that would mean that the algebra
of perturbative BPS states for Het on $T^d$ would be given by the tree level 
computation, at least perturbatively.

World-sheet arguments for the perturbative non-renormalization of 
massless $n$-point amplitudes, $0\leq n\leq 3$, have appeared in a
number of papers, starting with \marti. These first arguments overlooked
subtleties that arise when integrating out the fermionic moduli, since this
causes the supersymmetry current to develop unphysical poles \verlinde .
The residue at these poles can be written as a contribution coming from the 
boundary of moduli space, when the world-sheet Riemann surface degenerates 
into two Riemann surfaces connected by a long thin tube, creating an ambiguity
in the definition of the scattering amplitude. Those ambiguities were
analyzed extensively in \catop, whose results are consistent with the claim
that these amplitudes are perturbatively not renormalized. It would be 
interesting to check in detail whether the analysis of \catop\ carries on to 
arbitrary BPS states.

Non-perturbatively, we need an space-time approach. For massless states,
there is an argument due to Dine and Seiberg \ds, showing that the 
$0\leq n \leq 3$ massless scattering amplitudes with 16 supercharges are 
$exact$. However, since we don't have a space-time formalism dealing with 
arbitrary BPS states, we will just be able to comment on the more obvious 
sources of non-perturbative effects.

Instanton corrections could come from wrapped Euclidean heterotic 5-branes
but they can not contribute until we compactify on $T^6$. On the other hand, 
by Shenker's argument \shenker, we expect $e^{-{1\over g}}$ effects in every 
string theory, including heterotic theory (see \eva\ for such effects for 
heterotic strings). A way to rule out such corrections to the tree level 
result would be to argue that for heterotic on $T^6$, the BPS algebra is 
holomorphic in the complexified heterotic coupling constant $\tau$, since 
$e^{-{1\over g}}$ is not holomorphic in $\tau$. If so, by decompactifying 
dimensions of $T^6$, one could argue that these corrections are not present 
for higher dimensional toroidal compactifications either. 

In short, there is some evidence, at least for mutually local BPS states,
that the algebra of BPS states with 16 supercharges is independent of the 
coupling constant. From the heterotic side, this seems plausible at the 
perturbative level. However, currently we lack a framework to study this
issue non-perturbatively.

\newsec {BPS spectrum and algebra for Type IIA on $\IC^2/\IZ_N$}

Compactifications of type IIA theory on $\IC^2/\IZ_N$/ALE space are
 considerably simpler than on a  Calabi-Yau threefold, but they are a 
very good example to illustrate our proposals in previous 
sections. First we present the physical results (which can 
be obtained using various dualities) and then we 
analyze to what extent we can derive them from quiver theory.

\subsec {Physical results}

Consider IIA on $\IC^2/\Gamma$, where $\Gamma$ is a finite subgroup of 
$SU(2)$. It has associated with it 
a simply laced Dynkin diagram ${\cal G}$, an affine
algebra $\hat G$ and a Lie algebra $G$. At the orbifold point there are  
2-cycles shrunk to zero volume; their intersection matrix is the Cartan 
matrix of the algebra $G$. Recall that we can have a B field at the shrunk 
cycles \aspin. If the $B$ field is turned on, the 
conformal field theory is well behaved, whereas
if the $B$ field is zero, the conformal field theory breaks down and we have
enhanced gauge symmetry, as required by duality with heterotic string on
${\bf T}^4$. We will mostly focus on $\Gamma =\IZ _N$, which gives an affine
$\hat A_{N-1}$ algebra. 

The possible BPS D-branes are D2-branes wrapping some set of 2-cycles, 
possibly carrying in addition the charge of $n$ D0-branes. The mass formula
for a brane wrapped about the cycle $\alpha$ carrying $n$ units of D0 charge
is
\eqn\mass{
m^2={1 \over g_s^2}\biggl((B_\alpha)^2+(J_\alpha)^2 +n^2 
\biggr),}
where $B_{\alpha}$, $J_{\alpha}$ denote the integral of $B$ and $J$, 
respectively, on the $\alpha$-cycle \foot{As noted above, 
we are choosing $\zeta_{\IC}=0$.}. The mass formula just tells us 
the mass that a BPS state present at a site
of the charge lattice would have. We want to
determine which sites of the charge lattice are indeed occupied by single 
particle BPS states, and what is the degeneracy of states at each site. 
In the next section we will see that we can answer these questions 
at least semiclassically using the methods of \refs {\pistab,\dfr}. 

The charge lattice can be identified with the positive root lattice of 
the affine Lie algebra $\hat G$. If we choose a basis of simple roots 
$\alpha_0$, $\alpha_1, \cdots, \alpha_n$, where $n$ is the 
rank of $G$, this lattice is 
\eqn\posroot{
\Gamma_+ =\{k_1 \alpha_1 + \cdots + k_n \alpha_n + k_0 \alpha_0
 |\, k_i \ge 0 \}.}  
The simple roots $\alpha_1, \cdots, \alpha_n$ correspond to the exceptional 
divisors that one obtains after resolving the singularity. The null root 
is defined as $\delta = \sum_i m_i \alpha_i$, where $m_i$ are the usual 
labels of the affine Dynkin diagram. For $\widehat A_{N-1}$, one has 
$\delta=\alpha_1 + \cdots+ \alpha_0$, and the $B$-field at the perturbative
orbifold is given by the 
periods $B_{\alpha _i}=1/N$ for the simple roots \refs {\aspin,\md}. 
We will always consider the orbifold with the $B$ field turned on. Also
 notice that the integrals of $J$ on the cycles $\alpha_i$ are the FI terms: 
$J_{\alpha_i}=\zeta_i$. We can write the mass $m_{\alpha}$ for a BPS state 
with charge $\alpha$ as $m_{\alpha}=||Z_{\alpha}||/g_s$, where $Z_{\alpha}$ 
is the central charge given by:
\eqn\centch{
Z_{\alpha}=\zeta \cdot \alpha + i {e \cdot \alpha \over N}.}
In this equation, $e=(1,1, \cdots, 1)$ as in \valor, and we have denoted 
$\zeta\cdot \alpha =\sum_i k_i \zeta_i$, $e \cdot \alpha = \sum_i k_i$.

From all the possible states in $\Gamma_+$, we only expect a subset  
in the physical spectrum of BPS bound states of the theory. These states 
are the following:

1) Positive roots of $G$, $\alpha_+$:  
They correspond to fractional branes \refs {\md,\ddg,\dgm}, which 
are interpreted as D2 branes wrapping the 2-cycle
$\alpha_+$ in the homology lattice of the ALE space. 
For $\widehat A_{N-1}$ they have the form
\eqn\pos{
\alpha_{ij} =\alpha_i + \alpha_{i+1} + \cdots + \alpha_j,\,\,\,\,\ i<j.}
The central charge associated to a positive 
root is $Z_{\alpha_{ij}}= \zeta_i+ \cdots + \zeta_j + 
i|j-i+1|/N$, and 
we might have boundstates at threshold. We expect one bound state for each 
positive root, since if we go now to the orbifold with $B_{\alpha_+}=0$, 
duality with heterotic string on ${\bf T}^4$ requires the existence of 
these states \sd, and they can not disappear from the spectrum. 
They correspond 
to the W bosons of the enhanced gauge symmetry, and they have no degeneracy.

2) Null roots of $\hat G$. They have the form $n\delta$, with 
$n>0$. They correspond to a boundstate with $n$ 
units of D0 charge. The mass is given by $m_{n \delta}=n/g_s$, so for 
$n>1$, if they 
exist they are boundstates at threshold. Indeed, 
M/IIA duality requires their existence. A way to see this is to 
increase the coupling, so we are in $M$ theory on 
${\rm ALE}\times {\bf S}^1$: 
we have to get all the KK modes, and we expect  
these states to be present also at $g_s=0$. 

3) Roots of the form $\alpha_+ +n\delta$, with $n>0$: They 
correspond to the KK modes of the W bosons. They can be at threshold 
for special values of $\zeta$. Again, M/IIA duality requires their existence.

4) Roots of the form $-\alpha_+ +n\delta$, with $n>0$. For $n=1$, 
they are naturally interpreted as anti-D2 branes \dg. The states 
with $n>1$ correspond to the KK modes of these anti-D2 branes. 

Notice that the expected spectrum consists of the positive roots of 
the affine Lie algebra. (1), (3) and (4) correspond to the real roots:
\eqn\realro{
\Delta_+^{\rm real}(\hat G)=
\{\alpha_+ + n \delta,\,\, \alpha_+ \in \Delta_+(G),\,\, n\ge 0 \}
\cup  \{ -\alpha_+ + n \delta,\,\, \alpha_+ \in \Delta_+(G),\,\, n> 0 \},}
while (2) corresponds to the imaginary roots:
\eqn\imro{
\Delta_+^{\rm im}(\hat G)=\{ n \delta | \, n>0 \}.}

\subsec {The quiver theory.}

The quiver diagram corresponding to D-branes on $\IC^2/\IZ_N$ is constructed 
in terms of the affine Dynkin diagram $\widehat A_r$, where 
$r=N-1$ is the rank \dm. These are precisely the quivers that appear 
in the work of Kronheimer and Nakajima \refs{\kron,\kn}. The quiver 
diagram that one obtains is the following: 
take the corresponding affine Dynkin diagram, and denote the nodes by 
$V_i$, $i=0,1, \cdots, r$. These nodes are complex vector spaces of 
dimension $n_i$, so $V_i \simeq \IC^{n_i}$, and the gauge group 
is $G=\prod_i U(n_i)$, with complexified gauge group $G_{\IC}= 
\prod_i {\rm Gl}(n_i, \IC)$. For every node there are two
arrows $B_{i,i+1}$, $B_{i, i-1}$ in opposite directions. 
The structure is the following:
\eqn\anquiver{
\diagram
&  & &  V_0&    &  & \\
 &  &\ldTo(3,2)^{\scriptstyle{B_{0,1}}}\,\,\ \ruTo(3,2)_{
\scriptstyle{B_{1,0}}}\,\,\,\,\, & 
  &  \,\,\,\,\,\  \luTo(3,2)_{\scriptstyle{B_{r,0}}}\,\,\  
 \rdTo(3,2)^{\scriptstyle{B_{0,r}}}& &  \\
V_1 &\pile{\rTo^{\scriptstyle{B_{1,2}}} \\ \lTo_{\scriptstyle{B_{2,1}}}} 
&V_2&
\pile{\rTo^{{\scriptstyle{B_{2,3}}}} \\ \lTo_{\scriptstyle{B_{3,2}}}}& 
\cdots &  
\pile{\rTo^{\scriptstyle{B_{r-1,r}}} \\ 
\lTo_{\scriptstyle{B_{r,r-1}}}} &V_{r}\\ 
\enddiagram}
 As we discussed above, there are complex and real equations. The  
complex equations come from the superpotential of the 
${\cal N}=2$ theory, which in the Higgs phase imposes the 
following relations:
\eqn\bes{
B_{i+1,i}B_{i, i+1}=B_{i-1,i}B_{i,i-1}.}
The real equations read:
\eqn\real{
D_k=B_{k-1,k}B^{\dagger}_{k-1,k}-B^{\dagger}_{k,k-1}B_{k,k-1} + 
B_{k+1,k}B^{\dagger}_{k+1,k}-B^{\dagger}_{k,k+1}B_{k,k+1}=\theta_k.} 

It is clear that life gets considerably simpler if we are left only 
with \bes, so we will look for $\theta$-stable $G_{\IC}$-orbits of the 
solutions to the complex equations. Because of our previous remarks, each 
orbit will contain a solution to \real. The dimension vectors for the 
representations of this quiver are in one-to-one correspondence with the 
lattice $\Gamma_+$, after identifying 
the simple root $\alpha_i$ with the dimension vector whose entries 
are all zero except at the $i$-th place, where the entry is one: 
$n=(0,\cdots, 0,1,0, \cdots, 0)$ (we identify $0\equiv N$). This is the 
dimension vector of the simple representation $U_i$, which has a single 
$\IC$ at the $i$-th node, and all the arrows are set to zero. 

Notice that, for this quiver, one has a supersymmetric vacuum only for 
$n =\delta=(1, 1, \cdots, 1)$, which describes the D0-brane. The other 
possible dimension vectors $n$ correspond to fractional branes  
and combinations thereof. These fractional branes are interpreted 
as wrapped D2-branes (or anti D2-branes) with D0-charge, and 
are described by quasi-supersymmetric vacua: the D terms 
are not zero, but take a nonzero, constant value which 
breaks supersymmetry. This gives the relation between the $\theta$ 
parameters and the FI terms stated in \valor.
 
As a final remark, notice that the quiver \anquiver, 
in contrast to the quivers 
considered in \dfr, is ``nonchiral,'' in the sense that  
there are two arrows with opposite orientations in each vertex. This is 
of course a consequence of ${\cal N}=2$ supersymmetry. Nonchiral quivers 
can be regarded as a direct sum of two chiral quivers with opposite 
orientations, and they have been considered in some detail in \nakapaper. 

\subsec{The BPS spectrum from the quiver theory}
    
Our purpose now is to find the $\theta$-stable
representations of the quiver \anquiver, where $\theta$ is 
related to $\zeta$ by \valor,  
and see that we recover all the physical 
expectations explained above. 

We will denote the dimension vector of the representation by 
$n$. The first thing we want to know is the dimension of the moduli 
space of representations ${\cal M}_\theta(n)$. 
As shown in \nakapaper, this space contains as 
an open subset the $\theta$-stable representations, and this open set 
has the expected complex dimension 
\eqn\dim{
2-n^t \cdot C \cdot  n,}
where $C$ is the generalized Cartan matrix of the quiver 
(in our case, the Cartan matrix of the affine $\widehat {SU(N)}$). 
The first thing 
to do is to see which dimension vectors give a nonnegative expected 
dimension. The relevant result here is due to Kac \kactwo\ and says 
that, for any Kac-Moody algebra with generalized Cartan matrix $C$, and 
for any vector of nonnegative components $n$, 
$ n^t \cdot C \cdot  n =2$ if and only if $ n$ is a real root 
of the algebra, and $ n^t \cdot C \cdot  n \le 0$ if and only if 
$n$ is an imaginary root. We then see that the only possible dimension 
vectors giving a nonnegative dimension space of $\theta$-stable 
representations are the positive roots $\Delta_+$ of the Kac-Moody algebra. 
This is already in agreement with the physical results 
established above. 

The next step is to analyze each of the positive roots and 
find the $\theta$-stable representations, for different 
values of the Fayet-Iliopoulos parameters. This will prove 
the existence of bound states.       

1) {\it Positive roots of G}. For the simple roots $\alpha_i$, the 
representation is given by $U_i$, $i=0,\cdots, N-1$.  
Since they don't have any proper 
subrepresentation, they are $\theta$-stable everywhere in moduli space, in a
trivial sense. 

For positive but not simple roots (the rest of the gauge multiplet) the story
is far more interesting. Consider the simple case of a positive root given by
the sum of two adjacent simple roots, $\alpha_i + \alpha_{i+1}$, $i=0,1, 
\cdots, N-1$, and we identify $0 \equiv N$. This corresponds to a D2-brane
wrapping two adjacent $\IP^1$'s. The complex equation \bes\ is easily 
solved in this case, and gives $B_{i,i+1}=0$ or $B_{i+1,i}=0$. Up to 
complex isomorphism, this gives two possible representations, with two 
different subobjects. For $B_{i+1,i}=0$ one has:
\eqn\primero{
\diagram
\cdots& 0 & \pile{\lTo^{\scriptstyle{0}} \\ \rTo_{\scriptstyle 0}}& 
\IC &\pile{\lTo^{\scriptstyle{\simeq}} \\ \rTo_{\scriptstyle 0}} &\IC & 
\pile{\lTo^{\scriptstyle{0}} \\ \rTo_{\scriptstyle 0}}&0& \cdots  
 \\ & & &\uTo_{\scriptstyle \simeq}& &\uTo_{\scriptstyle 0}& & & \\
\cdots& 0& \pile{\lTo^{\scriptstyle{0}} \\ \rTo_{\scriptstyle 0}}&
 \IC& \pile{\lTo^{\scriptstyle{0}} \\ \rTo_{\scriptstyle 0}} & 0&  \pile{\lTo^{\scriptstyle{0}} \\ \rTo_{\scriptstyle 0}}&0& \cdots \\
\enddiagram}
while for $B_{i,i+1}=0$ one has:
\eqn\segundo{
\diagram
\cdots& 0 & \pile{\lTo^{\scriptstyle{0}} \\ \rTo_{\scriptstyle 0}}& 
\IC &\pile{\lTo^{\scriptstyle{0}} \\ \rTo_{\scriptstyle {\simeq}}} &\IC & 
\pile{\lTo^{\scriptstyle{0}} \\ \rTo_{\scriptstyle 0}}&0& \cdots  
 \\ & & &\uTo_{\scriptstyle 0}& &\uTo_{\scriptstyle \simeq}& & & \\
\cdots& 0& \pile{\lTo^{\scriptstyle{0}} \\ \rTo_{\scriptstyle 0}}&
 0& \pile{\lTo^{\scriptstyle{0}} \\ \rTo_{\scriptstyle 0}} & \IC& 
 \pile{\lTo^{\scriptstyle{0}} \\ \rTo_{\scriptstyle 0}}&0& \cdots \\
\enddiagram}

The meaning of these diagrams is as follows: the top row is the original
representation, with $0$ and $\simeq$ denoting the zero and the 
identity map, respectively. Physically, these two possibilities 
mean that the chiral fields 
$B_{i,i+1}$, $B_{i+1,i}$
have a zero or non-zero vev. When the vev is nonzero, 
we perform a complex gauge transformation to set it equal to the  
identity map. The bottom row is the subrepresentation. All the
maps from the bottom to the top row have to be injective, by definition of
subrepresentation. Checking that these diagrams commute is extremely easy.
 The representation in \primero\ is
 $\theta$-stable if $\theta_i >0$, and the second 
representation \segundo\ is stable if $\theta_i<0$. 
If $\theta_i=0$, both are 
semistable, and both are S-equivalent to 
the direct sum $U_i \oplus U_{i+1}$, 
which is the graded representation. In our general discussion we stressed
that stability allows us to bypass solving the D-flatness equations; in this 
case it is immediate to check the equivalence of the two approaches: 
$\theta$-stable representations are in one-to-one correspondence with the 
solutions to the real equations, which in this case are simply:
\eqn\realeq{
|B_{i+1,i}|^2-|B_{i,i+1}|^2=\theta _i.}
For $\theta _i>0$, we have $|B_{i+1,i}|=\sqrt {\theta _i}$ and $B_{i,i+1}=0$, 
and for $\theta _i<0$, we have $B_{i+1,i}=0$ and 
$|B_{i,i+1}|=\sqrt {-\theta_i}$, in agreement with the stability analysis. 
For $\theta _i=0$, the solution is $B_{i+1,i}=B_{i,i+1}=0$, {\it i.e.}  
the graded representation. 

Physically, the semistable point $\theta_i=0$ corresponds to a line 
of marginal stability when $\zeta_i = \zeta_{i+1}$ in the space of FI terms, 
as one can see using \valor. One can easily check, using \mass, that on this 
line
\eqn\lineone{
|| Z_{\alpha_i + \alpha_{i+1}}|| = ||Z_{\alpha_i}|| + ||Z_{\alpha_{i+1}}||,} 
in agreement with the structure of the graded representation $U_i \oplus 
U_{i+1}$. However, there is no decay of the state through this line. 
This is because 
at both sides of the line there is a stable representation with dimension 
$n=\alpha_i + \alpha_{i+1}$, as we have just seen. Therefore, in these 
models {\it there are lines 
of marginal stability which do not give a decay}. As expected, this is 
due to the fact that our quiver is nonchiral, which is in turn a 
consequence of having sixteen supercharges in the bulk.
Notice that the ``binding'' of the BPS states, in the quiver picture, 
corresponds very precisely to the maps between the nodes.  

For the more general positive roots $\alpha_i +\cdots + \alpha_j$ 
the story is very similar. There are in general $j-i$ lines of 
marginal stability in the $\zeta$-space, and there are no decays through 
them: the BPS state exists on both sides of the lines. 

2) {\it Null roots of $\hat G$}. The moduli space ${\cal M}_{\zeta}(\delta)$ 
has real dimension $4$, and it is the ALE space itself, as 
it follows from the results of \kron\kn. 
Notice that 
there are lines of marginal stability in the space of FI terms. For example, 
for ${\widehat {SU(3)}}$, the lines will be at $\zeta_1=0$, 
$\zeta_2=0$, and $\zeta_1+ \zeta_2=0$, as one can easily check by using the 
mass formula. At these lines, the representations will be only semistable. 
The moduli space ${\cal M}_{\zeta}(\delta)$ can be understood as a 
resolution of singularities in the following sense \nakapaper. For 
$n=\delta$, all the maps in \anquiver\ are just complex numbers. If one 
defines:
$$
x=B_{1,0}B_{2,1} \cdots B_{n, n-1} B_{0,n},$$
\eqn\maseq{
y=B_{n,0}B_{n-1,n} \cdots B_{1,2} B_{0,1}, \,\,\,\,\ z=B_{1,0} B_{0,1},}
then the complex equations \bes\ imply that $xy=z^{n+1}$. This is the 
equation that defines the quotient $\IC^2/\IZ_N$ as a subspace of $\IC^3$. 
Therefore, one has a map: 
\eqn\reso{
\pi: {\cal M}_{\zeta}(\delta) \rightarrow \IC^2/\IZ_N.}
It can be seen that this map is an isomorphism outside the inverse 
image of the origin $\pi^{-1}(0)$. This set is called the {\it 
exceptional set}, 
and it is given by $N-1$ $\IP^1$'s (the exceptional divisors). 
We will denote them by $\Sigma_i$, 
$i=1, \cdots, N-1$.

The BPS spectrum is given by the $L^2$-cohomology of the ALE space 
$H_{L^2}^{*}({\cal M}_{\zeta}(\delta))$, which is 
concentrated in the middle dimension \hitchin. A convenient choice of basis 
is given by the Poincar\'e duals to the $\Sigma_i$, $\omega_i=[\Sigma_i]$. 
These cohomology classes satisfy \kn\nakapaper:
\eqn\norm{
\int_{\Sigma_k} \omega_j =C_{jk},}
where $C_{jk}$ is the Cartan matrix of the Lie algebra. We then find that 
there are $N-1$ BPS states with charge $\delta$. 
   
When the dimension vector is $n \delta$, the moduli space is the $n$-th 
symmetric product of the ALE, and at a generic point, the 
automorphism group is $(\IC^*)^n$, so the 
representation is not Schur, and therefore it cannot be 
stable. We have to restrict to the small diagonal \hm, which 
is a copy of the ALE space itself. The 
representations in the small diagonal are only semistable, but 
one expects that the cohomology (which is given again by the 
two-forms $\omega_i$, $i=1, \cdots, N-1$) corresponds to 
bound states at threshold. Summarizing, for any imaginary root $n\delta$ 
with $n>0$, the BPS spectrum is given by $N-1$ states which are in 
one-to-one correspondence with the $L^2$-cohomology of the ALE space. 

3) {\it The rest of the roots}. The rest of the roots have the 
form $\pm \alpha_+ + n \delta$, $n>0$. The moduli space has dimension 
zero, so ${\cal M}_{\theta}(\pm \alpha_+ + n \delta)$ is a set of 
points, each of them giving one BPS state (the space could be 
empty). In the next subsection, we will analyze in detail 
a particular case in affine $SU(3)$ 
and see that it gives the right degeneracy: one 
$\theta$-stable representation for each value of $\theta$, giving 
a single BPS state. It also shows a rich structure of lines of 
marginal stability and will illustrate our discussion of 
graded representations.  

\subsec{A detailed case study}

We will now study in detail the representations with dimension 
$n=\alpha_1 + \delta$ for ${\widehat {SU (3)}}$. This will illustrate 
very well the richness as well as the computability of the 
procedure. The quiver representation has the diagram:
\eqn\alpdel{
\diagram
 & & \IC & &  \\
  &\ldTo^{\scriptstyle b}\,\,\ \ruTo_{\scriptstyle \tilde b} &
  & \rdTo_{\scriptstyle {\tilde c}}\,\,\   \luTo^{\scriptstyle { c}} \\
\IC^2 & & \pile{ \rTo^{\scriptstyle a} \\ \lTo_{\scriptstyle{\tilde a}}} 
& &\IC\\ 
\enddiagram}
where $a$, $\tilde a$, $b$ and $\tilde b$ are regarded as two-component 
vectors, and one has for example $a\cdot(z_1,z_2)=a_1 z_1 + a_2 z_2$, 
$\tilde a \cdot z =(\tilde a_1 z, \tilde a_2 z)$, and so on. The complex 
equations \bes\ can be written as follows:
$$a\cdot \tilde a =b \cdot \tilde b =c\tilde c,$$
\eqn\compcomp{
b_1\tilde b_1=a_1 \tilde a_1, \,\,\ b_2\tilde b_2=a_2 \tilde a_2, 
\,\,\ b_1\tilde b_2=\tilde a_1  a_2, \,\,\ 
b_2\tilde b_1=\tilde a_2  a_1.}
and we want to find the $\theta$-stable representations, for arbitrary 
values of $\theta$. Since $\theta$-stable representations are Schur, a 
good starting point is to look for Schur representations. 

From the definition, we have to consider all the possible endomorphism of the 
generic representation \alpdel. These are given by $\lambda_1$, $\lambda_2 
\in \IC^*$, acting on the $\IC$ nodes, and a matrix $S\in {\rm Gl}(2, \IC)$ 
acting on $\IC^2$. Commutativity of all the diagrams gives the equations:
$$
(\lambda_1 -\lambda_2)c=(\lambda_1-\lambda_2) \tilde c=0, $$ 
\eqn\computa{
S^t \cdot a =\lambda_1 a, \,\,\ S\cdot \tilde a =\lambda_1 \tilde a, \,\,\ 
S \cdot b = \lambda_2 b, \,\,\ S^t \cdot \tilde b = \lambda_2 \tilde b.}
Schur representations are such that they force $\lambda_1=\lambda_2=\lambda$ 
and $S= \lambda \cdot {\bf 1}_{2 \times 2}$. A careful analysis of these 
equations shows that, for every value of $\theta$, there is only 
one Schur representation (up to complex 
isomorphism) which is $\theta$-stable and solves the complex equations. 
These solutions are as follows: 

a) For $\theta_1>0$, $\theta_1+\theta_2>0$ (region I in figure 3), 
the stable
representation is:
\eqn\alpdelone{
\diagram
 & & \IC & &  \\
  &\ldTo^{\scriptstyle i_2}\,\,\ \ruTo_{\scriptstyle 0} &
  & \rdTo_{\scriptstyle {\simeq}}\,\,\   \luTo^{\scriptstyle { 0}} \\
\IC^2 & & \pile{ \rTo^{\scriptstyle 0} \\ \lTo_{\scriptstyle{ i_1}}} 
& &\IC\\ 
\enddiagram}
where $i_p$ denotes the inclusion in the $p$-th factor. 
The subrepresentations have dimensions $(2,0,0)$, $(2,1,0)$ and $(1,1,0)$ 
(in the notation $n=(n_1,n_2, n_0)$). 
When $\theta_1>0$ but $\theta_1+\theta_2<0$ (region II), the $\theta$-stable 
representation is like \alpdelone, but with $c$ and $\tilde c$ exchanged. 
The subrepresentations have dimensions $(2,0,0)$, $(1,0,1)$ and $(2,0,1)$. 

b) For $\theta_1<0$, and $\theta_1 + \theta_2>0$ (region III), the $\theta$-stable 
representation is 
\ \eqn\alpdeltwo{
\diagram
 & & \IC & &  \\
  &\ldTo^{\scriptstyle 0}\,\,\ \ruTo_{\scriptstyle \pi_2} &
  & \rdTo_{\scriptstyle {\simeq}}\,\,\   \luTo^{\scriptstyle 0} \\
\IC^2 & & \pile{ \rTo^{\scriptstyle \pi_1} \\ \lTo_{\scriptstyle{0}}} 
& &\IC\\ 
\enddiagram}
where $\pi_p$ denote the projection onto the $p$-th factor. The 
subrepresentations of \alpdeltwo\ have dimensions $(0,1,0)$, 
$(0,1,1)$, $(1,1,0)$ and $(1,1,1)$. Finally, for $\theta_1<0$ and 
$\theta_1 + \theta_2<0$ (region IV in figure 3), 
the $\theta$-stable representation is like \alpdeltwo, but again with 
$c$ and $\tilde c$ exchanged. The subrepresentations have 
dimensions $(0,0,1)$, $(0,1,1)$, 
$(2,0,1)$ and $(1,1,1)$. 

\ifig\regionsfig{Lines of marginal stability for $\alpha_1+ \delta$}
{\epsfxsize2.0in\epsfbox{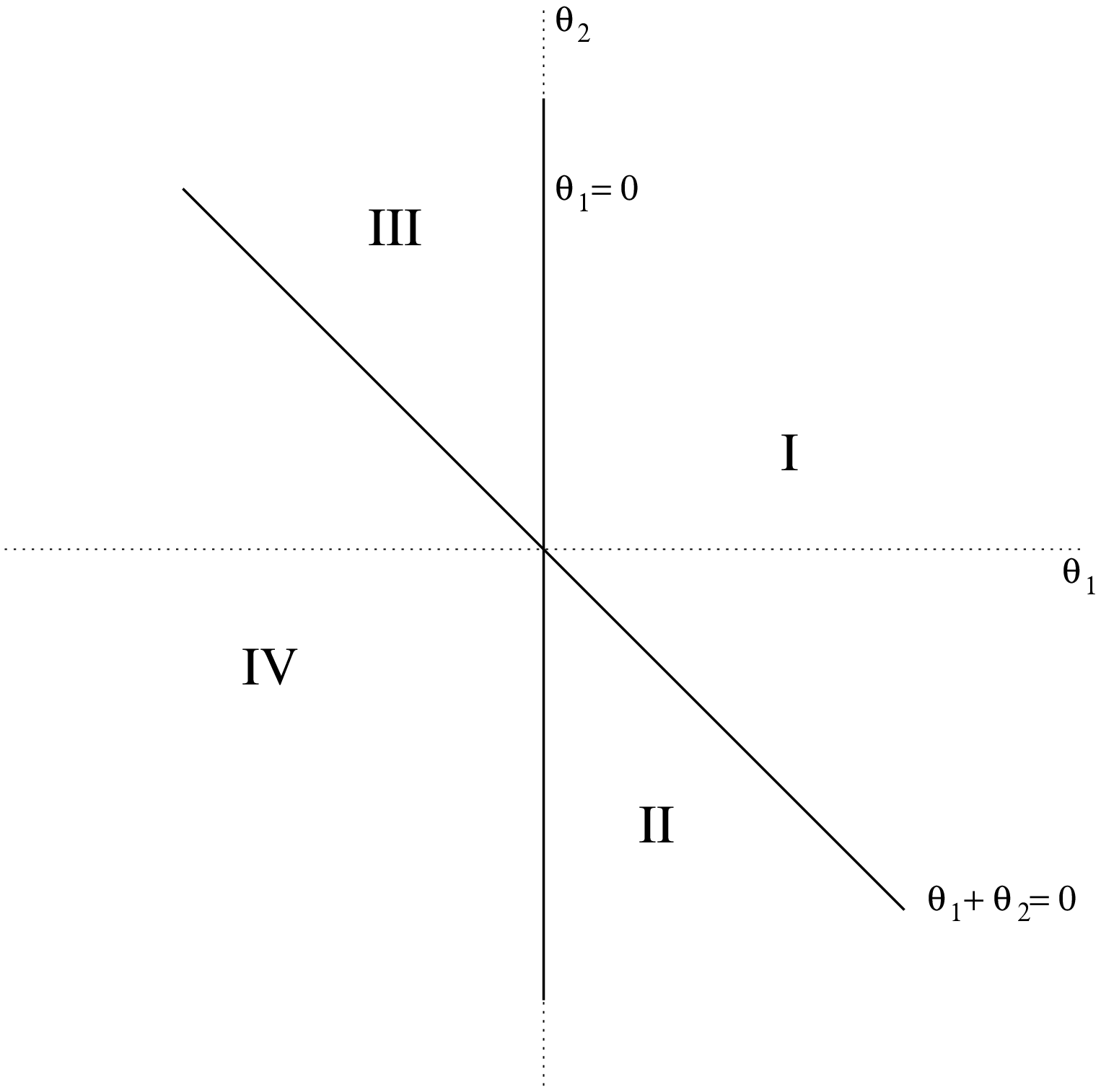}}

We see that there are two lines of marginal stability, at 
$\theta_1=0$ and at $\theta_1 + \theta_2 =0$, but again no decay takes place 
through them, since $\theta$-stable representations exist on both sides 
of the lines. Let's analyze in more detail what happens on these 
lines. The representations that we 
have described are only semistable there, and they will have a 
nontrivial Jordan-H\"older filtration. Since we have listed all the 
subobjects, we can construct the filtrations very easily, for the 
different lines of marginal stability. Consider first 
the half-line $\theta_1=0$, $\theta_2>0$, which separates region I from 
region III. The representation \alpdeltwo\ 
becomes semistable and 
has the following filtration: 
\eqn\jhone{
\diagram
&& & & \IC & &  \\
&&  &\ldTo^{\scriptstyle 0}\,\,\ \ruTo_{\scriptstyle \pi_2} & 
\uTo
  & \rdTo_{\scriptstyle {\simeq}}\,\,\   \luTo^{\scriptstyle { 0}} \\
R&&\IC^2 & \pile{ \rTo^{\scriptstyle{\pi_1}} \\ \lTo_{\scriptstyle{ 0}}} 
&\HonV& &\IC\\ 
&&&  & \vLine^{\scriptstyle \simeq}&
 & \\
\cup &&\uTo^{\scriptstyle{i_1}}& & \IC & &\uTo_{\scriptstyle \simeq} \\
&& &\ldTo^{\scriptstyle 0}\,\,\ \ruTo_{\scriptstyle 0}& \uTo & 
 \rdTo_{\scriptstyle {\simeq}}\,\,\   \luTo^{\scriptstyle { 0}} \\
R_2&&\IC & \pile{ \rTo^{\scriptstyle \simeq} \\ \lTo_{\scriptstyle{ 0}}} 
&\HonV& &\IC\\
&& &  & \vLine^{\scriptstyle \simeq}&
 &\\
\cup&&\uTo^{\scriptstyle 0}& & \IC & & \uTo_{\scriptstyle \simeq}\\
&& &\ldTo^{\scriptstyle 0}\,\,\ \ruTo_{\scriptstyle 0}&   & 
 \rdTo_{\scriptstyle {\simeq}}\,\,\   \luTo^{\scriptstyle { 0}} \\
R_1 && 0 & \pile{ \rTo^{\scriptstyle 0} \\ \lTo_{\scriptstyle{0}}} 
& & &\IC\\
\enddiagram} 
Notice that all the representations in the filtration satisfy 
$\sum_v n_v \theta_v=0$ on the line $\theta_1=0$. The last representation 
is stable, since its only subobject has dimension $(0,1,0)$. The two 
quotients involved in the filtration are isomorphic 
to the simple representation $U_1$, therefore they are stable. 
This shows that the above is an admissible Jordan-H\"older filtration.     
The graded representation is:
\eqn\grone{
\diagram
 & & \IC & &  \\
&\ldTo^{\scriptstyle 0}\,\,\ \ruTo_{\scriptstyle 0} &
  & \rdTo_{\scriptstyle {\simeq}}\,\,\   \luTo^{\scriptstyle { 0}} \\
0 & & \pile{ \rTo^{\scriptstyle 0} \\ \lTo_{\scriptstyle{0}}} 
& &\IC\\ 
\enddiagram \,\,\ {\displaystyle \oplus}\,\,\ 
{\displaystyle 2}\,\,\
\diagram
 & & 0 & &  \\
&\ldTo^{\scriptstyle 0}\,\,\ \ruTo_{\scriptstyle 0} &
  & \rdTo_{\scriptstyle {0}}\,\,\   \luTo^{\scriptstyle { 0}} \\
\IC & & \pile{ \rTo^{\scriptstyle 0} \\ \lTo_{\scriptstyle{0}}} 
& &0\\ 
\enddiagram}
The $\theta$-stable representation on the other side of the line (region I)
is \alpdelone. It also becomes semistable on the line, 
and its graded representation is also \grone. Therefore, the 
$\theta$-stable representations on both sides of the line of marginal 
stability become S-equivalent on it. 

As a further example, let's consider the line $\theta_1 + \theta_2 =0$, 
and $\theta_1<0$, which separates region I from region II. 
The representation \alpdeltwo\ becomes 
again semistable, but in this case it has a very 
different Jordan-H\"older filtration:
\eqn\jhtwo{ 
\diagram
& & & & \IC & &  \\
& &  &\ldTo^{\scriptstyle 0}\,\,\ \ruTo_{\scriptstyle \pi_2} & 
\uTo
  & \rdTo_{\scriptstyle {\simeq}}\,\,\   \luTo^{\scriptstyle { 0}} \\
R & & \IC^2 & \pile{ \rTo^{\scriptstyle{\pi_1}} \\ \lTo_{\scriptstyle{ 0}}} 
&\HonV& &\IC\\ 
& & &  & \vLine^{\scriptstyle 0}&
 & \\
\cup & & \uTo^{\scriptstyle{i_1}}& & 0 & &\uTo_{\scriptstyle \simeq} \\
& & &\ldTo^{\scriptstyle 0}\,\,\ \ruTo_{\scriptstyle 0}&   & 
 \rdTo_{\scriptstyle {0}}\,\,\   \luTo^{\scriptstyle { 0}} \\
R_1 & & \IC & &\pile{ \rTo^{\scriptstyle \simeq} \\ \lTo_{\scriptstyle{ 0}}} 
& &\IC\\
\enddiagram}
The graded representation is in this case:
\eqn\grtwo{
\diagram
 & & 0 & &  \\
&\ldTo^{\scriptstyle 0}\,\,\ \ruTo_{\scriptstyle 0} &
  & \rdTo_{\scriptstyle {0}}\,\,\   \luTo^{\scriptstyle { 0}} \\
\IC & & \pile{ \rTo^{\scriptstyle \simeq} \\ \lTo_{\scriptstyle{0}}} 
& &\IC\\ 
\enddiagram \,\,\,\,\ {\displaystyle \oplus}\,\,\,\,\ 
\diagram
 & & \IC & &  \\
&\ldTo^{\scriptstyle 0}\,\,\ \ruTo_{\scriptstyle \simeq} &
  & \rdTo_{\scriptstyle {0}}\,\,\   \luTo^{\scriptstyle { 0}} \\
\IC & & \pile{ \rTo^{\scriptstyle 0} \\ \lTo_{\scriptstyle{0}}} 
& &0\\ 
\enddiagram}
Again, the $\theta$-stable representation on the other 
side of the line has the same graded representation on it, and it becomes 
S-equivalent to \alpdeltwo. 

The above graded representations correspond nicely to our physical 
expectations based on the central charges. Using \valor, one finds 
that $\theta_1$, $\theta_2$ are related to the FI parameters as 
follows:
\eqn\relatio{
\theta_1 = {3 \over 4} \zeta_1, \,\,\,\,\ \theta_2=\zeta_2 - {1 \over 4} 
\zeta_1,}
Therefore the lines of marginal stability are $\zeta_1=0$, 
$\zeta_1 + 2\zeta_2=0$. When $\zeta_1=0$, the central charge satisfies:
\eqn\firstz{
||Z_{\alpha_1 + \delta}||=2||Z_{\alpha_1}|| + ||Z_{-\alpha_1 + \delta}||,}
and the decomposition is precisely the one in \grone. Also, for 
$\zeta_1 + 2\zeta_2=0$ one finds
\eqn\secondz{
||Z_{\alpha_1 + \delta}||=||Z_{\alpha_1 + \alpha_2}|| + 
||Z_{-\alpha_2 + \delta}||,}
as dictated by \grtwo. This shows very clearly that the decompositions given 
by the graded representations are the minimal ones, since the 
subobjects appearing in the direct sum are strictly stable on the line.

\subsec {Computation of the BPS algebra}

The spectrum that we have found matches perfectly 
with the generators of the positive part of the 
Kac-Moody algebra: the $r$ states with charges $n\delta$ (which 
have the charges of $n$-D0 branes) give the
generators $H_n^i$, $i=1\dots r$, $n>0$. The state of a D2 
wrapping $\alpha_+$ with
$n$ D0-branes, which has charge $\alpha_+ + n\delta$, 
 corresponds to $E^{\alpha_+}_n$, with $n\ge 0$. 
The anti-D2 branes together with their KK modes give the generators 
$E^{-\alpha_+}_n$, $n>0$. These generators give a subalgebra of the 
full Kac-Moody algebra $\widehat A_r$, with the following relations (we 
follow the conventions of \kacbook):
\eqn\posalgebra{
\eqalign{
& [H_m^i, H_n^j]=0, \,\,\,\,\,\ m,n >0,\cr
& [H^i_m, E^{\alpha}_n]=\alpha^i E^{\alpha}_{m+n}, \,\,\,\,\,\ m>0,\,\ 
n \ge 0, \,\, \alpha \in \Delta_+,\cr
&[E_m^\alpha, E_n^{-\alpha}]=-\alpha \cdot H_{m+n},\,\,\,\,\,\ m\ge 0,\,\ 
n > 0, \,\, \alpha \in \Delta_+,\cr
&[E_m^\alpha, E_n^{\beta}]=\epsilon(\alpha,\beta)E_{m+n}^{\alpha + \beta}
,\,\,\,\,\,\ \alpha,\, \beta,\, \alpha + \beta   \in \Delta_+,\cr
}}
where, in the last commutator, $m,n$ are nonnegative or positive 
when $\alpha$, $\beta$ are positive or negative, respectively. The $\epsilon 
(\alpha, \beta)$ are the $\IZ_2$ cocycles and 
take values $\pm 1$. Notice that there are many 
consistent choices of $\epsilon(\alpha, \beta)$, which turn out to 
be related to choices of orientation of the quiver diagram 
\kacbook\ringeltwo. 
Finally, the roots $\alpha$ in the l.h.s. of \posalgebra\ 
are understood as vectors in the Dynkin basis with components 
$\alpha^i$, and one has for 
example for the simple roots $(\alpha_k)^i=C_{ki}$, so that 
$\alpha_k \cdot H_n = C_{ki}H^i_n$. 

Due to heterotic/type IIA duality, the algebra of D2-D0 BPS 
states that we have considered should be the subalgebra \posalgebra\ 
of the full Kac-Moody algebra. This has been explicitly checked in \hm\ 
by a vertex operator computation in the heterotic side. 
We will now compute the BPS algebra in a few cases starting from the 
definition in terms of representations of quivers.

Consider first the positive roots of the simple Lie algebra $A_r$, and the 
simple case of $n_1=\alpha_i$, $n_2=\alpha_{i+1}$ and $n_3 =\alpha_i + 
\alpha_{i+1}$. The moduli spaces of representations ${\cal M}_\zeta (n_i)$ 
are nonempty, consist of a single point, and this point is a stable 
representation away from the marginal stability line $\zeta_i =\zeta_{i+1}$, 
as we saw in section 4. The correspondence variety is also a 
point. When $\zeta_i > \zeta_{i+1}$, this point corresponds to the 
short exact sequence:
\eqn\bpsone{
\diagram
R_2&& \cdots& 0 & \pile{\lTo^{\scriptstyle{0}} \\ \rTo_{\scriptstyle 0}}& 
0&\pile{\lTo^{\scriptstyle{0}} \\ \rTo_{\scriptstyle 0}} &\IC & 
\pile{\lTo^{\scriptstyle{0}} \\ \rTo_{\scriptstyle 0}}&0& \cdots  
 \\ \uTo& &  & & &\uTo_{\scriptstyle 0}& &\uTo_{\scriptstyle \simeq}& & & \\
R_3 & & \cdots& 0 & \pile{\lTo^{\scriptstyle{0}} \\ \rTo_{\scriptstyle 0}}& 
\IC &\pile{\lTo^{\scriptstyle{\simeq}} \\ \rTo_{\scriptstyle 0}} &\IC & 
\pile{\lTo^{\scriptstyle{0}} \\ \rTo_{\scriptstyle 0}}&0& \cdots  
 \\\uTo && & & &\uTo_{\scriptstyle \simeq}& &\uTo_{\scriptstyle 0}& & & \\
R_1 & & \cdots& 0& \pile{\lTo^{\scriptstyle{0}} \\ \rTo_{\scriptstyle 0}}&
 \IC& \pile{\lTo^{\scriptstyle{0}} \\ \rTo_{\scriptstyle 0}} & 0&  \pile{\lTo^{\scriptstyle{0}} \\ \rTo_{\scriptstyle 0}}&0& \cdots \\
\enddiagram}     
while for $\zeta_i < \zeta_{i+1}$, one has the sequence with the roles of 
$U_i$ and $U_{i+1}$ inverted. Recall that, when the correspondence variety 
is a point, one has simply to count short exact sequences. We then find:
\eqn\firstcomputa{
[E_0^{\alpha_i}, E_0^{\alpha_{i+1}}]={\rm sgn}(\zeta_i-\zeta_{i+1})
E_0^{\alpha_i + \alpha_{i+1}}.}
Notice that the sign $\epsilon(\alpha_i, \alpha_{i+1})={\rm sgn}(\zeta_i-
\zeta_{i+1})$ depends on the 
region of the moduli space we are considering. This suggests that 
the different regions of stability in the $\zeta$-space are 
related to choices of orientation of the underlying diagram, and 
in fact one can check that this is the case in concrete examples. 

 The computation of the 
structure constants for the other positive roots of $A_r$ 
follows the same lines, 
and it's rather easy. In all cases one finds the last relation of 
\posalgebra, and the sign depends again on the precise values on the 
region of the moduli space of FI parameters that we are sitting in. 
We would like to make two remarks: first, this computation is essentially 
the computation of the Ringel-Hall algebra associated to the $A_r$ 
Dynkin diagram (see for example the recent paper \frenkel). Second, it's 
remarkable that such a simple computation already captures an important 
part of the physics associated to the BPS states in this compactification, 
namely the emergence of Lie algebra structures. 

Let's now come to the part of the algebra involving positive roots 
of the full Kac-Moody algebra. As in the previous subsection, we will make 
a sample computation in the case of ${\widehat {SU(3)}}$. This 
computation will involve nontrivial correspondence varieties of positive 
dimension, and prove the power of the quiver approach in performing 
actual computations. 

Consider first the case $n_1=\alpha_1$, $n_2 =\delta$, 
$n_3=\alpha_1 + \delta$. We will choose a $\zeta$ such that 
$\zeta_1<0$, $\zeta_2<0$, and $\zeta_1+ 2\zeta_2<0$ (this choice is made 
for convenience, since in that case the arguments in \nakapaper\ make 
possible to identify very quickly the correspondence variety). In this 
region, 
the stable representative of $\alpha_1 + \delta$ is given by \alpdeltwo, 
but with the arrows on the right exchanged ({\it i.e.}, in the notation of 
subsection 4.4, $c$ is exchanged with $\tilde c$). In order to construct 
the short exact sequence, we need $n_1$ or $n_2$ to be dimension vectors 
of subrepresentations. In this case, there are no subrepresentations with 
dimension $(1,0,0)$, therefore ${\cal C}(n_1,n_2;n_3)=0$. It turns out 
that there is a whole family of 
embeddings of representations with dimension vector $(1,1,1)$ into 
the $\theta$-stable representation with dimension $\alpha_1 + \delta$. The 
short exact sequence that one obtains is: 
\eqn\bpstwod{\diagram
& &  & & 0 & &  \\
& &   &\ldTo^{\scriptstyle 0}\,\,\ \ruTo_{\scriptstyle 0} & 
\uTo
  & \rdTo_{\scriptstyle {0}}\,\,\   \luTo^{\scriptstyle { 0}} \\
R_1 & & \IC & \pile{ \rTo^{\scriptstyle{0}} \\ \lTo_{\scriptstyle 0 }} 
&\HonV& &0\\ 
& & &  & \vLine^{\scriptstyle 0}&
 & \\
\uTo & &\uTo^{\scriptstyle{\pi_{(\lambda,\mu)}}}& & \IC & &\uTo_{\scriptstyle 0} \\
& & &\ldTo^{\scriptstyle 0}\,\,\ \ruTo_{\scriptstyle {\pi_2}}& \uTo & 
 \rdTo_{\scriptstyle {0}}\,\,\   \luTo^{\scriptstyle { \simeq}} \\
R_3 & & \IC^2 & \pile{ \rTo^{\scriptstyle{\pi_1}} \\ \lTo_{\scriptstyle{ 0}}} 
&\HonV& &\IC\\
& &&  & \vLine^{\scriptstyle \simeq}&
 &\\
\uTo & & \uTo^{\scriptstyle{i_{(\lambda, \mu)}}}& & \IC & & \uTo_{\scriptstyle \simeq}\\
& &  &\ldTo^{\scriptstyle 0}\,\,\ \ruTo_{\scriptstyle \mu}&   & 
 \rdTo_{\scriptstyle {0}}\,\,\   \luTo^{\scriptstyle { \simeq}} \\
R_2 & & \IC & &\pile{ \rTo^{\scriptstyle {\lambda}} \\ \lTo_{\scriptstyle{0}}} 
&  & \IC\\
\enddiagram}
In this diagram, $\mu$ and $\lambda$ are complex numbers, and 
the maps $i_{(\lambda, \mu)}$ and $\pi_{(\lambda,\mu)}$ are given by 
\eqn\mapas{
i_{(\lambda, \mu)}(z)=(\lambda\, z, \mu \, z), \,\,\,\,\,\,\
\pi_{(\lambda,\mu)}(z_1,z_2)=\mu z_1 -\lambda z_2,}
in such a way that $\pi_{(\lambda,\mu)}i_{(\lambda, \mu)}=0$, as required 
by exactness. The space of short exact sequences is then parametrized by the 
pairs $(\lambda, \mu)$. But we have to take into account two things: 
first of all, $(\lambda, \mu) \not=(0,0)$, since in that case the 
first representation in the short exact sequences would be decomposable, 
contradicting stability. On the other hand, one has to quotient by the 
complexified gauge group acting on the representation. We then consider 
endomorphisms $\alpha_i: \IC \rightarrow \IC$, where $\alpha_i \in 
\IC^*$. $i=1,2,0$, and they act by multiplication on the nodes of 
the representation. The space of complexified gauge transformations actually 
has dimension two, so we have to get rid of one extra 
degree of freedom. Since we want to study the effect of gauge 
transformations on the maps $(\lambda, \mu)$, we fix this 
extra degree by requiring that the transformed 
representations have the same form as in \bpstwod, {\it i.e.} that they leave 
the arrows on the right unchanged in the representations with dimension 
$(1,1,1)$. This imposes $\alpha_0=\alpha_2$. Therefore, the action of the 
complexified gauge group is: 
\eqn\act{
(\lambda, \mu) \rightarrow ( \alpha \lambda, \alpha \mu),}
where $\alpha=\alpha_1^{-1} \alpha_2 \in \IC^*$. The space that 
parameterizes the short exact sequences is nothing but the projective space 
$\IP^1$. We then find:
\eqn\corone{
{\cal C}(n_2, n_1;n_3) =\IP^1.} 
Another way to see this is to notice that the space of short 
exact sequences \bpstwod\ 
is in one-to-one correspondence with the space of embeddings 
of $\IC$ in $\IC^2$ up to complex isomorphisms 
(the maps that we have denoted by $i_{(\lambda, \mu)}$). 
This space is nothing but the Grassmannian $G(1,2) \simeq 
\IP^1$. Notice that, from this point of view, 
$(\lambda,\mu)=(0,0)$ is forbidden 
by injectivity of $i_{(\lambda, \mu)}$. 
 
In order to compute the product, we have to be more explicit about 
this space. First of all, notice that this $\IP^1$ can be identified 
in a natural way with a subspace of the space of $\theta$-stable 
representations with dimension $(1,1,1)$, which is the ALE space. Moreover, 
the $\IP^1$ is contained in the exceptional set, as one can see using 
\maseq. We can go further and identify this $\IP^1$ with one of the 
$\Sigma_i$'s. To do this, we use Theorem 5.10 in \nakapaper, which gives 
a very concrete recipe to identify representations in $\pi^{-1}(0)$ with 
the exceptional $\IP^1$'s. The recipe is the following: consider the 
representation $R$ in question, and construct a filtration:
\eqn\filtrasin{
0\subset R_1 \subset \cdots \subset R_{p} \subset R,}
in such a way that all the quotients $R_i/R_{i-1}$ are isomorphic 
to simple representations $U_i$. The existence of such a decomposition 
is guaranteed by the fact that $R \in \pi^{-1}(0)$. Moreover, 
$R \in \Sigma_j$ if and only if $R/R_{p} \simeq U_j$. Notice that 
the above is not a Jordan-H\"older filtration, since it doesn't take into 
account $\theta$-stability \foot{However, it can be regarded 
as the Jordan-H\"older filtration of $R$ when $\theta=0$.}. It corresponds 
in fact to the decomposition of $R$ into simple pieces. In our example, 
the construction of the filtration is rather easy. It has 
$p=2$, $R_1=U_0$ and $R_2$ is 
\eqn\reptwo{   
\diagram
 & & \IC & &  \\
&\ldTo^{\scriptstyle 0}\,\,\ \ruTo_{\scriptstyle 0} &
  & \rdTo_{\scriptstyle {0}}\,\,\   \luTo^{\scriptstyle { \simeq}} \\
0 & & \pile{ \rTo^{\scriptstyle 0} \\ \lTo_{\scriptstyle{0}}} 
& &\IC\\ 
\enddiagram}
Therefore, $R/R_2 =U_1$ and the correspondence variety \corone\ is in fact 
$\Sigma_1$. Finally, using \norm, one finds 
\eqn\bpsconstwo{
[H^i_1, E^{\alpha_1}_0]=C_{i1}E_1^{\alpha_1},}
and since $\alpha_1^i=C_{i1}$, we find perfect agreement with \posalgebra. 
Using the above information, one can also compute the BPS product 
for $n_1=\alpha_1$, $n_2 =-\alpha_1 + \delta$, $n_3=\delta$, and obtain
\eqn\mas{
[E_0^{\alpha_1},E_1^{-\alpha_1}]=-\sum_i C_{1i} H^i_1.} 
This is again in agreement with 
\posalgebra. 

In conclusion, we see that representations 
of quivers give a framework for computing BPS algebras 
in an efficient way, since the use of linear algebra 
makes possible to give precise descriptions of the correspondence 
variety of \hm.

\newsec{Conclusions} 

In this paper we have used the notions of stability and S-equivalence 
to clarify the structure of D-brane moduli spaces in different regions 
of the K\"ahler moduli space. In particular, we have seen that 
one can give a detailed recipe to obtain decay products on lines of 
marginal stability by using graded objects. We have also shown that 
the algebra of BPS states and the correspondence conjecture of \hm\ 
can be extended to the orbifold point, 
and formulated in terms of representations of quivers. This gives 
a very useful framework to study and compute BPS algebras. All these 
ideas have been tested in the rather simple example of type IIA 
theory on an ALE space, where we can compare the results obtained 
from our constructions with expectations based on the use of 
various dualities. 
 
An obvious direction for future work is to study
lines of marginal stability and algebras of BPS states for Calabi-Yau 
orbifolds. The immediate candidate is $\IC^3/\IZ_3$, since its BPS 
spectrum near the orbifold has been studied in \dfr. The 
adjacency matrix of the corresponding quiver gives a generalized 
Cartan matrix of the hyperbolic type, and 
it is likely that the BPS algebra of fractional branes 
will turn out to be related to the hyperbolic 
Kac-Moody algebra associated to the Cartan matrix. 
Also, as mentioned in \dfr\ and 
fully worked out in \ddm, quiver techniques can be applied to Gepner points, 
giving a handle on compact Calabi-Yau manifolds. We also expect 
the algebra of BPS 
states to be related to the Kac-Moody algebra defined 
by the corresponding Cartan matrix. This would be give a very 
interesting generalization of the Nakajima construction \nakalg\ 
providing a geometrical setting for non-affine Kac-Moody algebras.  

A different approach to some of these questions has been proposed 
recently in \refs {\mirrorvafa,\mirrord,\lerche}, where Landau-Ginzburg (LG) 
duals are constructed for 
many compact and noncompact Calabi-Yau manifolds, and explicit descriptions 
for the D-branes on both sides are provided. It would be 
very interesting to see what are the mirror descriptions of BPS spectra, 
lines 
of marginal stability, decay products, and BPS algebras, using the LG 
approach. This may be helpful as well to obtain the right category of 
objects at arbitrary points of moduli space. 

\vfill    
\eject

\bigskip
\centerline{\bf Acknowledgments}
\bigskip

We would like to thank T. Banks, J. Gomis, J. Harvey, J. Maldacena, 
H. Nakajima, C. R\"omeslberger and A. Schofield for useful discussions 
and correspondence. We are specially indebted to M. Douglas and G. Moore 
for very fruitful discussions and comments on the manuscript.

The diagrams in this paper were prepared with the package 
diagrams.tex, created by Paul Taylor. We would like to 
thank him for making it available to the scientific community 
with no charge.  
  
This work has been supported by DOE grant DE-FG02-96ER40959.

\vfill    
\eject

\appendix{A}{$\Pi$-stability and abelian categories} 

In this appendix we prove that the subcategory of $\Pi$-semistable D-branes
with a fixed grading $\varphi$ at a given point in moduli space $u$, is 
abelian. We use arguments valid for small difference of gradings.

First we give a couple of informal definitions. A category is just a set of 
objects and maps between them. A category is additive when there is some 
notion of sum in it, such that the sum of two objects is also an object in
the same category. Finally, a category is abelian if it is additive and
has the notion of kernel and cokernel for the maps in the category.

Let us call ${\cal A}$ the category of all branes satisfying the holomorphic
constraint. As we explained in the text, we assume ${\cal A}$ is abelian.
Consider the subcategory $A_\varphi(u)$  
of D-branes with fixed grading $\varphi$, that are $\Pi$-semistable at the 
point in moduli space $u$. Recall that the D-brane $E$ is $\Pi$-semistable if
$\varphi(E)\geq \varphi (E')$
for all the subobjects $E'$ of $E$ at $u$.

First let's prove that $A_\varphi(u)$ is additive. For that we need to define
the notion of sum, and check that it takes two elements of $A_\varphi(u)$ to
another element in $A_\varphi(u)$. The notion of sum is given by exact 
sequences. If we have three D-branes $E$, $F$ and $G$ forming the 
exact sequence
\eqn\shorts{
0 \longrightarrow E {\buildrel \phi \over \longrightarrow} F 
 {\buildrel \psi \over \longrightarrow}
G \longrightarrow 0,
}
and $E$ and $G$ are in $A_\varphi(u)$, we want to prove that $F$ is also 
in $A_\varphi(u)$, 
$i.e.$ that $\varphi(F)\geq \varphi (F')$ for all the subobjects $F'$ of $F$ 
at $u$. Take such an $F'$. Let $G'=\psi (F')$, and define $E'$ by 
the short exact sequence 
\eqn\shortstwo{
0\rightarrow E'\rightarrow F'\rightarrow G'\rightarrow 0,}
in other words, $\phi(E')={\rm Ker}\, \phi |_{F'}$. $E'$ and $G'$ are 
subobjects of $E$ and $G$, respectively, and by assumption 
$\varphi(E')\leq \varphi(E)$ and $\varphi(G')\le \varphi(G)$. 
Here is where we restrict to small difference of gradings. This just means
that the central charges are just complex numbers, but in this case the
phases satisfy a convexity condition 
\eqn\gradconve{
\varphi(F')=x\varphi(E')+(1-x)\varphi (G'),}
for some $0\leq x\leq 1$. Using that $\varphi(E')\leq \varphi(E)$ and 
$\varphi(G')\leq \varphi(G)$ we obtain $\varphi (F')\leq \varphi(F)$.

Now we want to prove that $A_\varphi(u)$ is abelian. This requires that every
map between objects has a kernel and a cokernel in the category. Consider
the map $f:E\rightarrow F$, with $E$ and $F$ in $A_\varphi(u)$. First we argue
that $\varphi({\rm Im}\ f)=\varphi$. On one hand, ${\rm Im}\ f$ 
is a subobject of $F$, so
$\varphi({\rm Im}\ f)\leq \varphi$ by assumption. On the other hand, 
we have the 
exact sequence $0\rightarrow {\rm Ker}\ \varphi\rightarrow E\rightarrow 
{\rm Im}\ \varphi
\rightarrow 0$, so $\varphi ({\rm Im}\ f)\geq \varphi$, and our claim 
follows. 
From the exact sequence we just wrote it follows that $\varphi ({\rm Ker}\ f)=
\varphi$. Furthermore, ${\rm Ker}\ f$ has to be $\Pi$-semistable, 
because every 
subobject of ${\rm Ker}\ f$ is also subobject of $E$, so a subobject 
destabilizing ${\rm Ker}\ f$ would also destabilize $E$. A dual of this 
argument shows that ${\rm Coker}\ f$ is also $\Pi$-semistable.

\listrefs
\end